\documentclass[aps,preprint,floatfix,nofootinbib,groupedaddress,superscriptaddress]{revtex4-1}
\usepackage{amsmath}
\usepackage{amssymb,graphics,graphicx,times,bm,multirow,color,setspace}

\begin{document}

\title{Characterization and quantification of the role of coherence in ultrafast
quantum biological experiments using quantum master equations, atomistic simulations, and quantum process tomography}

\author{Patrick Rebentrost}
\thanks{ Contributed equally to this work. These three authors are ordered in the sequence that their main scientific contribution is emphasized in the text.}
\author{Sangwoo Shim}
\thanks{ Contributed equally to this work. These three authors are ordered in the sequence that their main scientific contribution is emphasized in the text.}
\author{Joel Yuen-Zhou}
\thanks{ Contributed equally to this work. These three authors are ordered in the sequence that their main scientific contribution is emphasized in the text.}
\author{Al\'an Aspuru-Guzik}
\email{aspuru@chemistry.harvard.edu}
\affiliation{Department of Chemistry and Chemical Biology, Harvard University, 12 Oxford
St., Cambridge, MA 02138}

\begin{abstract}
  Long-lived electronic coherences in various photosynthetic complexes
  at cryogenic and room temperature have generated vigorous efforts
  both in theory and experiment to understand their origins and
  explore their potential role to biological function. The ultrafast
  signals resulting from the experiments that show evidence for these
  coherences result from many contributions to the molecular
  polarization. Quantum process tomography (QPT) is a technique whose
  goal is that of obtaining the time-evolution of all the density
  matrix elements based on a designed set of experiments with
  different preparation and measurements. The QPT procedure was
  conceived in the context of quantum information processing to
  characterize and understand general quantum evolution of
  controllable quantum systems, for example while carrying out quantum
  computational tasks. We introduce our QPT method for ultrafast
  experiments, and as an illustrative example, apply it to a
  simulation of a two-chromophore subsystem of the Fenna-Matthews-Olson
  photosynthetic
  complex, which was recently shown to have long-lived quantum
  coherences. Our Fenna-Matthews-Olson model is constructed using an atomistic approach
  to extract relevant parameters for the simulation of photosynthetic
  complexes that consists of a quantum mechanics/molecular mechanics
  approach combined with molecular dynamics and the use of
  state-of-the-art quantum master equations. We provide a
  set of methods that allow for quantifying the role of quantum
  coherence, dephasing, relaxation and other elementary processes in
  energy transfer efficiency in photosynthetic complexes, based on the
  information obtained from the atomistic simulations, or, using QPT,
  directly from the experiment. 
  The ultimate
  goal of the combination of this diverse set of methodologies is to
  provide a reliable way of quantifying the role of long-lived quantum
  coherences and obtain atomistic insight of their causes.

\end{abstract}
\date{ \today }

\maketitle

\section{Introduction}
\label{sec:1}

The initial step in photosynthesis is highly efficient excitonic transport of the energy captured from photons to a reaction center \cite{Blankenship2002}. In most plants and photosynthetic organisms this process occurs in light-harvesting complexes which are interacting chlorophyll molecules embedded in a solvent and a protein environment \cite{Cheng2009}. Several recent experiments show that excitonic coherence can persist for several hundreds of femtoseconds even at physiological temperature \cite{Engel2007,Lee2007,Panitchayangkoon2010,Collini2010}. These experiments suggest the hypothesis that quantum coherence is biologically relevant for photosynthesis. The results have motivated a sizeable amount of recent theoretical work regarding the reasons for the long-lived coherences and their role to the function.

The focus of many studies is on the theoretical models employed. In this context,
it is essential to be as realistic an possible and employ the least
amount of approximations. Most of the currently-employed methods
involve a master equation for the reduced excitonic density operator
where the vibrational degrees of freedom (phonons) of the protein and
solvent are averaged out. Amongst these simple methods are the Haken-Strobl model and
Redfield theory as employed in Refs. \cite{Rebentrost2009b,Plenio2008} and \cite{Mohseni2008} respectively.
 To interpolate between the usual weak and strong
exciton-phonon coupling limits, Ishizaki and Fleming developed a
hierarchical equation of motion (HEOM) theory which takes into account
non-equilibrium molecular reorganization effects
\cite{Ishizaki2009}. Jang et al. perform a second order time-convolutionless
expansion after a small polaron transformation to include strong
coupling effects \cite{Jang2008}.Another set of studies focuses on the role of quantum coherence and the phonon environmentin terms of transport efficiency or entanglement. It was
shown that the transport efficiency is enhanced by the interaction or interplay of the quantum evolution with the phononic environment \cite{Rebentrost2009b,Plenio2008,Mohseni2008,Wu2010}.
Entanglement between molecules is found to persist for long times \cite{Sarovar2010,Caruso2010a,Fassioli2010}.

The ongoing effort can be summarized with two equally important questions:
What are the microscopic reasons for the persistence of quantum
coherence and what is the relevance of the quantum effect to the biological functionality
of the organism under study? In this work, we summarize the recent efforts from our group to approach the problem from several angles. Firstly, we investigate the role of coherences in the exciton transfer process of the Fenna-Matthews-Olson (FMO) complex. We quantify the amount and the contribution of coherence to the efficient energy transfer process. Secondly, we present our quantum mechanics/molecular mechanics (QM/MM) approach to obtain information about the system at the atomistic level, such as detailed bath dynamics and spectral densities. Finally, we propose a spectroscopic tool that allows for obtaining directly the information of the quantum process via our recent theoretical proposal for the quantum process tomography technique to the ultrafast regime.

\section{The Role of Quantum Coherence}
\label{sec:coherence}
In this section, we discuss the question about the relevance of  quantum effects to the biological
function. A negative answer to this question would
mean that a particular effect, while being quantum, is not leading to
any improvement in the functionality of a biological system, and therefore would be a byproduct
of the spatial and temporal scales and physical properties of the
problem. For example, in energy transfer (ET) quantum coherence could arise from
the closely packed arrangement of the chromophores in a protein scaffold but it could, in
principle, represent a byproduct of that arrangement and not a relevant
feature. Another example, it may be true that the human eye can detect
a single photon, but it is not clear if this quantum effect
is relevant to the biological function, which usually operates at much larger photon fluxes.
If, on the other hand, the above yes-no question of the relevance is answered positively
for a particular effect in a biological system, it would present a major step
towards establishing the relevance or importance of a quantum biological phenomenon. A natural follow-up
questions is: How important \textit{quantitatively} is a particular quantum
effect?

Both of these questions should preferably be studied by experimental
means. An experiment would have to be designed in a way that tests for the
biological relevance of quantum coherence. Possible experiments could involve
quantum measurements on mutated samples. In
the FMO complex that acts as a molecular ET wire the efficiency of the
transport event is most likely a good quantifier for biological
function. One would need a way to experimentally quantify this efficiency
and extract the relevance of quantum coherence to the efficiency. This can
be hard in practice. Yet, as we will discuss in this work, quantum process
tomography is able to obtain detailed information about quantum coherence
and the phonon environment and might thus lead to progress in this area.

In the case when experimental access to an observable that involves the
biological relevance is hard or impossible, a theoretical treatment can provide
insight. 
It is illustrative to analyze a model of the particular biological process in terms of a quantifier for the success of the
process. An example is the aforementioned efficiency of energy transport. In
bird vision, the quantum yield of a chemical reaction is a relevant measure \cite{Cai2010}. Once a
detailed model and a success criterion is established, one needs to quantify
the contribution of quantum coherence to the success criterion. For this
step, one can proceed in two distinct pathways. The first pathway is a
comparison to a classical reference point; the success criterion is computed
for the actual system/model and a classical reference model that does not include quantum correlations.
The difference of these two values is attributed to quantum mechanics and can be considered
the quantum mechanical contribution to the success of the process.
For example, the energy transfer dynamics of a sophisticated quantum mechanical model such as
\cite{Ishizaki2009} could be compared to a semi-classical F\"{o}rster treatment that leads to a hopping
description.
In general, this comparison strategy has the drawback that one has to invoke a classical, and in some cases very artificial, model.

Our work has been mainly concentrated on a second theoretical pathway in answering the relevance question, which overcomes
this issue. It is based on just the quantum mechanical model and
the success quantifier. No other, for example classical, model is invoked.
The actual model will contain dynamical processes that are quantum
coherent and others that are incoherent. The non-trivial task is to deconstruct how the
various processes contribute to the performance criterion. This can be done
by decomposing the performance criterion into a sum of contributions, each
associated with a particular process. The terms in this sum related to
quantum mechanical processes will then give a theoretical answer to the
overall relevance of the particular process and will quantify this
relevance.
This line of thought was developed and discussed in Ref. \cite{Rebentrost2009a} for
energy transfer in the FMO complex and provided insight into both questions
"Is a quantum effect relevant?" and "If yes, how much?", at least from a theoretical standpoint within the approximations of the model under consideration.
In this section, we extend this idea to include the effect of the initial conditions
and compare the results to a total integrated coherence, or concurrence, measure. We utilize  secular Redfield theory and the hierarchy equation of motion approach.


The Hamiltonian describing a single exciton is given by:
\begin{equation}
H_e=\sum_m (\epsilon_m + \lambda) |m\rangle \langle m| +\sum_{m<n} J_{mn} \left(|m\rangle \langle n|+|n\rangle \langle m| \right).
\end{equation}
where the site energies $\epsilon_m$, and couplings $J_{mn}$ are usually obtained from detailed quantum chemistry studies and/or fitting of experimental spectra. The reorganization energy $\lambda$, which we assume to be the same for each site, is the energy difference of the non-equilibrium phonon state after Franck-Condon excitation and the excited-state equilibrium phonon state. The set of states $|m\rangle$ is called the site basis and the set of states $|\alpha\rangle$ with $H_e |\alpha\rangle = E_{\alpha} |\alpha\rangle$ is called the exciton basis.
We now briefly introduce the secular Redfield master equation in the weak exciton-phonon (or system-bath) coupling limit and the non-perturbative hierarchy equation of motion approach. In both approaches, the dynamics of a single exciton is governed by a master equation, which is schematically given by:
\begin{equation} \label{eqMasterEquation}
\frac{\partial}{\partial t} \rho(t)= \mathcal{M} \rho(t) = \left( \mathcal{M}_{\rm H} + \mathcal{M}_{\rm decoherence} +
\mathcal{M}_{\rm trap}+\mathcal{M}_{\rm loss} \right ) \rho(t).
\end{equation}
The master equation consists of the superoperator $\mathcal{M}$, which is divided into several components. First, coherent evolution with the excitonic Hamiltonian $H_{e}$ is described by the superoperator $\mathcal{M}_{\rm H} = -i [H_{e},\cdot ]$. In addition, decoherence due to the interaction with the phonon bath is incorporated by $\mathcal{M}_{\rm decoherence}$. $\mathcal{M}_{\rm decoherence}$ depends on the spectral density, which models the coupling strengths of the phonon modes to the system. Finally, one has the processes for trapping to a reaction center $\mathcal{M}_{\rm trap}$ and exciton loss $\mathcal{M}_{\rm loss}$ due to spontaneous emission. Associated with these processes are the trapping rate $\kappa$ and the loss rate $\Gamma$.
Details about the trapping and exciton loss processes can be found in \cite{Rebentrost2009a,Olaya-Castro2008}.

The secular Redfield theory is valid in the regime of weak system-bath coupling. The superoperator $\mathcal{M}_{\rm decoherence}$ is of Lindblad form with Lindblad operators for relaxation in the exciton basis and for dephasing of excitonic superpositions. The relaxation rates depend on the spectral density evaluated at the particular excitonic transition frequencies, satisfy detailed balance, and depend on temperature through the Bose-Einstein distribution. The dephasing rates are linear in temperature. We use the same Ohmic spectral density as in \cite{Ishizaki2009}, i.e.
$J(\omega)=2\lambda \gamma \omega /\pi(\omega^2 + \gamma ^2)$, where $1/\gamma$ is the bath correlation time. For $1/\gamma =50$ fs, this spectral density shows only modest differences to the spectral density used in \cite{Rebentrost2009a}. Further details about the Lindblad model can be found in \cite{Rebentrost2009a}. 

The hierarchy equation of motion approach \cite{Ishizaki2009} consistently interpolates between weak and strong system bath coupling. The assumption that the fluctuations are Gaussian makes the second-order cumulant expansion exact. The resulting equation of motion can be expressed as an infinite hierarchy of system, i.e. $\rho(t)$, and connected auxiliary density operators $\{\sigma_i\}$, arranged in tiers. For numerical simulation, "far-away" tiers in the hierarchy are truncated in a sensible manner. The hierarchy equation of motion can also be written as in Eq. (\ref{eqMasterEquation}) when we make the replacement $\rho(t) \to (\rho(t), \sigma_1, \sigma_2, \cdots)$ and use the hierarchical structure discussed in \cite{Ishizaki2009} for the decoherence superoperator $\mathcal{M}_{\rm decoherence}$. For simulations of the Fenna-Matthews-Olson complex, we use the scaled hierarchy approach developed in \cite{Shi2009}. It was shown recently that four tiers of auxiliary density operators are enough for accurate room temperature simulations \cite{Zhu2010}, which enables the rapid computation of efficiency and total coherences. The trapping and exciton loss processes are naturally extended to the auxiliary systems.

In our previous work \cite{Rebentrost2009a}, we developed a method to quantify the role of quantum coherence to the transfer efficiency. The energy transfer efficiency (ETE) is given by the integrated probability of leaving the system from the sites that are connected to the trap instead to being lost to the environment. That is, $\eta=\int_0^{\infty} dt {\rm Tr} \{ \mathcal{M}_{\rm trap} \rho(t) \}$. It was shown that the ETE can be partitioned into $\eta=\eta_{\rm H} + \eta_{\rm decoherence}$, where the efficiency due to the coherent dynamics with the excitonic Hamiltonian is given by:
\begin{equation}\label{eqContributionCoherent}
\eta_{\rm H}= {\rm Tr} \{ \mathcal{M}_{\rm trap} (\mathcal{M}_{\rm trap}+\mathcal{M}_{\rm loss})^{-1} \mathcal{M}_{\rm H} \mathcal{M}^{-1} \rho(0) \}.
\end{equation}
The ETE contribution $\eta_{\rm decoherence}$ involves $\mathcal{M}_{\rm decoherence}$, i.e. $\eta_{\rm decoherence}= {\rm Tr} \{ \mathcal{M}_{\rm trap} (\mathcal{M}_{\rm trap}+\mathcal{M}_{\rm loss})^{-1} \mathcal{M}_{\rm decoherence} \mathcal{M}^{-1} \rho(0) \}$. In this work, we extend our ETE contribution method to quantify the role of the initial state to the ETE. We obtain a separation of the coherent contribution, $\eta_{\rm H}= \eta_{\rm init}+\eta_{\rm dyn}$, where the efficiency $\eta_{\rm init}$ can be ascribed to the initial state. The $\eta_{\rm dyn}$ is defined by $\eta_{\rm dyn}=\eta_{\rm H}-\eta_{\rm init}$ and can be interpreted as dynamical part of the coherence contribution arising during the time evolution. For the computation of $\eta_{\rm init}$, we note that one can always express the ensemble described by the system density matrix as $\rho(t) = p_{\rm init}(t) |\psi_{\rm init}(t)\rangle \langle \psi_{\rm init}(t) | + \sum_{k} p_{k}(t) \rho_k(t)$. Here, $p_{\rm init}(t)$ is the probability of the quantum system being in the (Hamiltonian time-evolved) initial state $|\psi_{\rm init}(t)\rangle$, where $p_{\rm init}(0)=1$. The $p_{k}(t)$ are the probabilities of being in some other ensemble state $\rho_k(t)$, where $p_{\rm init}(t) + \sum_k p_{k}(t)=1$. The probability $p_{\rm init}(t)$ is modified by the interaction with the environment and readily computed for Markovian Lindblad dynamics by considering the damped no-jump evolution due to the decoherence superoperator $\mathcal{M}_{\rm decoherence}$ \cite{Olaya-Castro2008,Piilo2008,Rebentrost2009}. Therefore, we can compute the efficiency pertaining to the initial state by $\eta_{\rm init}=\int_0^{\infty} dt {\rm Tr} \{ \mathcal{M}_{\rm trap} p_{\rm init}(t) |\psi_{\rm init}(t)\rangle \langle \psi_{\rm init}(t) |\}$. Together with Equation (\ref{eqContributionCoherent}), this obtains the desired separation $\eta_{\rm H}= \eta_{\rm init}+\eta_{\rm dyn}$.

Additionally, we employ another measure for the role of coherence by straightforwardly integrating over time all the coherence elements of the density matrix. That is:
\begin{equation}\label{eqCoherenceMeasure}
C(\lambda)=\sum_{m\neq n} \int_{0}^{\infty }dt~|\left\langle m\right\vert \rho \left( t\right)
\left\vert n\right\rangle |.
\end{equation}%
We normalize with respect to the case of coherent evolution at $\lambda=0.0/$cm, i.e. $\tilde{C}(\lambda)=C(\lambda)/C(0)$. Based on the discussion in ~\cite{Sarovar2010}, the quantity $\tilde{C}$ can be considered as the (normalized) integrated entanglement (concurrence) that is present before the exciton is trapped in the reaction center or lost to the environment. We note that the total coherence measure $\tilde{C}$ is similar in spirit to a measure of the first kind discussed above. This is because the normalization essentially performs a comparison of the actual model at a certain $\lambda$ with an artificial model at $\lambda=0$. (For the numerical evalutation, the integral in Eq. (\ref{eqCoherenceMeasure}) is computed until ${\rm Tr}\{ \rho(t)\} \le 10^{-3} $.)

\begin{figure}[tbph]
\includegraphics[scale=0.7]{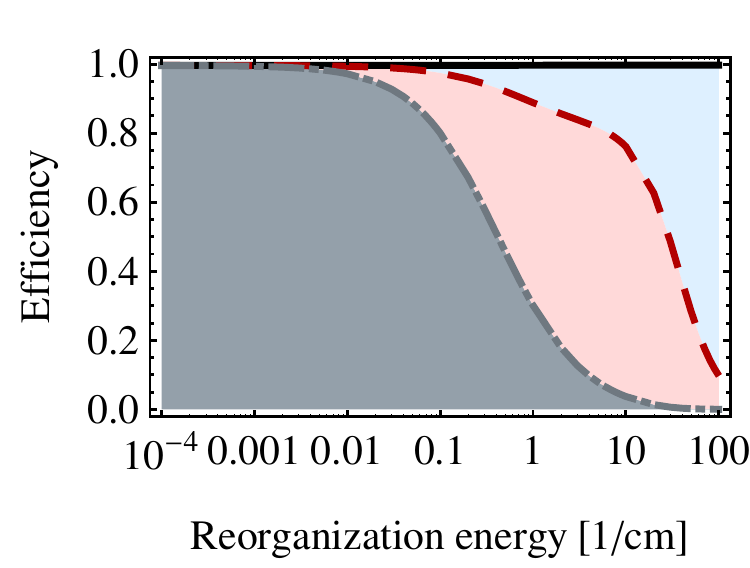}
\includegraphics[scale=0.7]{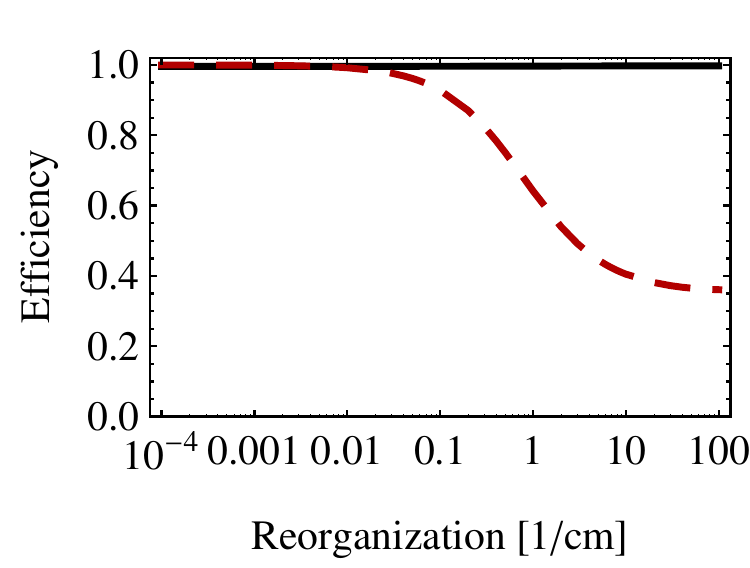}
\includegraphics[scale=0.7]{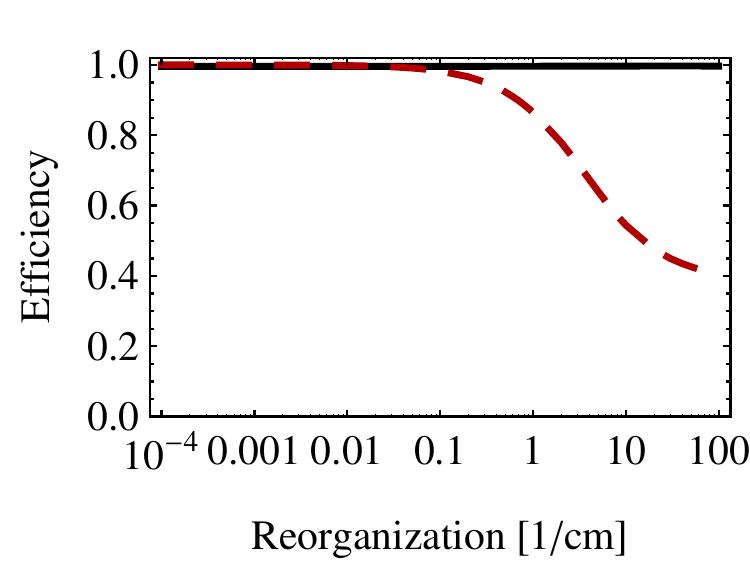}
\caption{
(Left panel) Efficiency $\eta$ (solid black) and contributions of initial state $\eta_{\rm init}$ (dash-dotted gray) and coherent evolution $\eta_{\rm init}+\eta_{\rm dyn}$ (dashed red) for a dimer that is based on the strongly coupled sites 1 and 2 of the Fenna-Matthews-Olson complex using the secular Redfield model. The initial state is at site 1 and the target is site 2. At a physiological value of around $\lambda=35/$cm, one finds $\eta_{\rm init} = 0.0$ and $\eta_{\rm dyn}= 0.43$.
(Center panel) Efficiency and integrated coherence $\tilde{C}$ for the dimer with the secular Redfield approach. At $\lambda=35/$cm there is $\tilde{C}=0.37$.
(Right panel) Same quantities as in the center panel for the dimer using the hierarchy equation of motion approach with 15 tiers of auxiliary systems. At $\lambda=35/$cm, one finds $\tilde{C}=0.44$. The parameters are $1/\protect\kappa=1$ ps, $1/\protect\Gamma=1$ ns, and $1/\protect\gamma=50$ fs for all panels.
}
\label{figContributionDimer}
\end{figure}

\begin{figure}[tbph]
\includegraphics[scale=0.7]{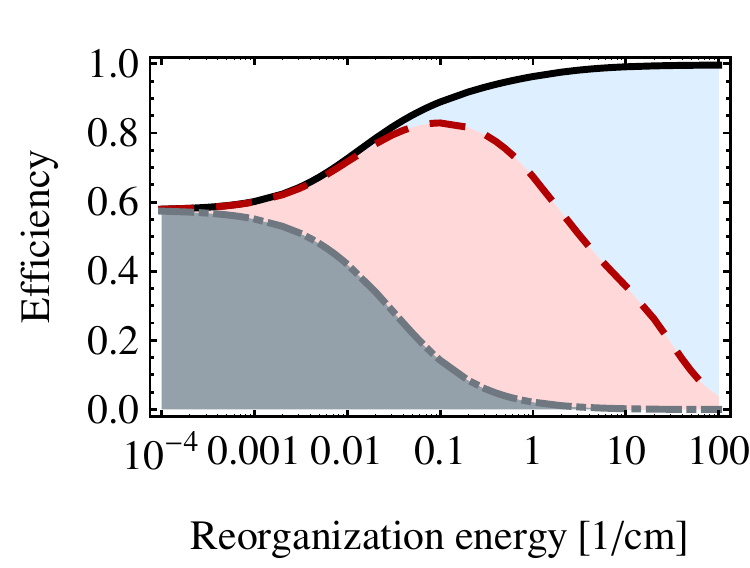}
\includegraphics[scale=0.7]{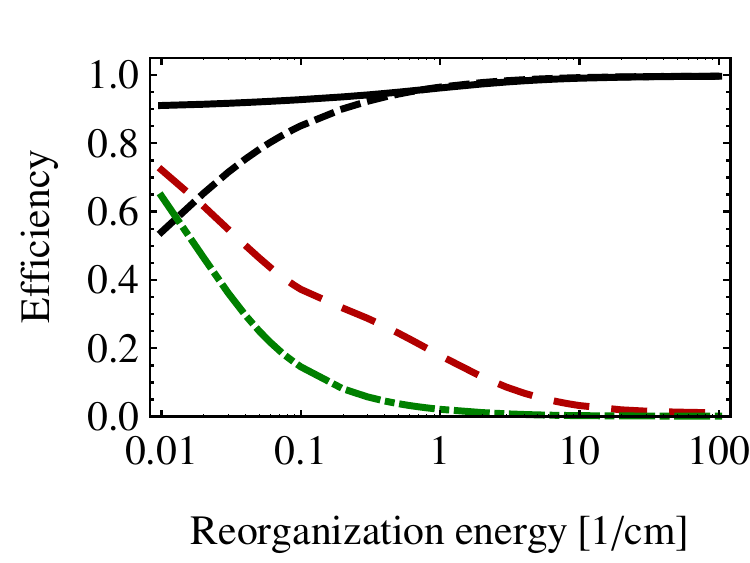}
\includegraphics[scale=0.7]{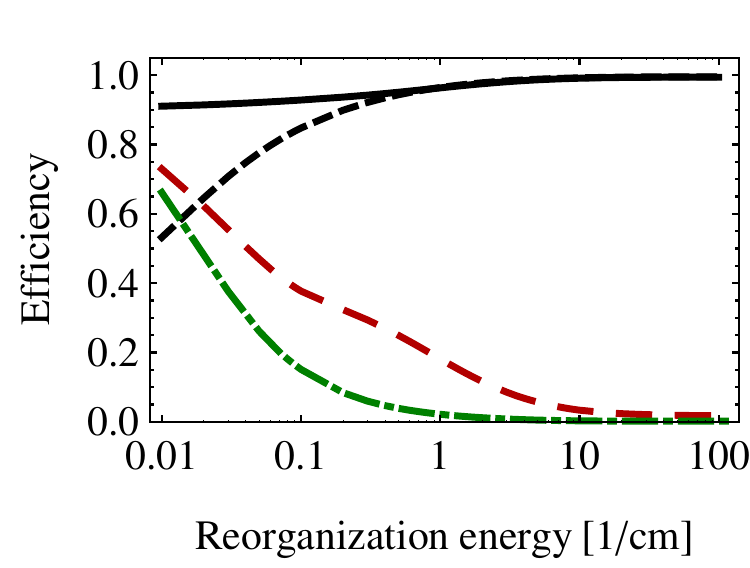}
\caption{
(Left panel) Efficiency $\eta$ (solid black) and contributions of initial state $\eta_{\rm init}$ (dash-dotted gray) and coherent evolution $\eta_{\rm init}+\eta_{\rm dyn}$ (dashed red) for the Fenna-Matthews-Olson complex using the secular Redfield model. The initial state is a classical mixture of site 1 and 6 and the target site for trapping is site 3. The actual system has a reorganization energy of around $\lambda=35/$cm, where $\eta_{\rm init} = 0.0$ and $\eta_{\rm dyn} = 0.17$.
(Center panel) Efficiency for initial site 1 (solid black) and initial site 6 (dashed black) and integrated coherence $\tilde{C}$ for initial site 1 (dashed red) and initial site 6 (dash-dotted green) for the Fenna-Matthews-Olson complex with the secular Redfield approach.
At $\lambda=35/$cm there is $\tilde{C}=0.0151$ (inital site 1) and $\tilde{C}=0.0017$ (initial site 6).
(Right panel) Same quantities as in the center panel for the FMO complex using the scaled hierarchy equation of motion approach with four tiers of auxiliary systems. At $\lambda=35/$cm, one finds $\tilde{C}=0.020$ (inital site 1) and $\tilde{C}=0.0022$ (initial site 6).
The parameters are $1/\protect\kappa=1$ ps, $1/\protect\Gamma=1$ ns, and $1/\protect\gamma=50$ fs for all plots. 
}
\label{figContributionFMO}
\end{figure}

In Fig. \ref{figContributionDimer}, we present the two measures of coherence for a dimer system. For the dimer, we take the sites 1 and 2 of the FMO complex with $\epsilon_1=0/$cm, $\epsilon_2=120/$cm, and $J=-87.7/$cm, see \cite{Adolphs2006}, and room temperature. This system will also be the focus of the following sections on the atomistic detail simulations and quantum process tomography. Here, for studying the role of quantum coherence, we assume that the task is defined by the exciton initially being at the lower energy site 1 and the target site being site 2. In the left panel of Fig. \ref{figContributionDimer} we show the efficiency $\eta$, the contribution $\eta_{\rm H}$ from Eq. (\ref{eqContributionCoherent}), and $\eta_{\rm init}$ for the secular Redfield model. In the present small system, environment-assisted transport is relatively unimportant, with the efficiency as a function of the reorganziation energy being close to unity everywhere. The underlying contributions show a transition from a regime dominated by coherent evolution to a regime dominated by incoherent Lindblad jumps. At $\lambda=35/$cm, we find $\eta_{\rm init}=0\%$ and $\eta_{\rm H}=43\%$. In Fig. \ref{figContributionDimer} (center panel), we find that the total coherence measure $\tilde{C}$ for the dimer is around $0.37$ for $\lambda=35/$cm. 
In Fig. \ref{figContributionDimer} (right panel), the total coherence is plotted for the dimer in the hierarchy equation of motion approach. We use 15 tiers of auxiliary systems. At $\lambda=35/$cm, we find $\tilde{C}=0.44$; because of the sluggish, non-equilibrium bath there is more coherence than in the secular Redfield model.

In Fig. \ref{figContributionFMO} (left panel), we present the coherent, decoherent, and initial state contribution to the ETE for the Fenna-Matthews-Olson complex as a function of the reorganization energy for the secular Redfield model at room temperature. We use the Hamiltonian given in \cite{Adolphs2006} and the contribution measures given in Equation (\ref{eqContributionCoherent}) and by $\eta_{\rm init}$. The initial state is a classical mixture of site 1 and 6. For small reorganization energy, the efficiency is around $\eta=60$\% and for larger reorganization energies we observe environment-assisted quantum transport (ENAQT) ~\cite{Rebentrost2009b}, with the efficiency rising up to almost $\eta=100$\% for the physiological value of $\lambda=35/$cm. The contributions measures $\eta_{\rm dyn}$ and $\eta_{\rm init}$ reveal the underlying dynamics. The quantum dynamical contribution $\eta_{\rm dyn}$ is around $17$\% at $\lambda=35/$cm \footnote{In Ref. \cite{Rebentrost2009a}, we found the value  $\eta_{\rm H}=10$\% for a different Hamiltonian and a different spectral density.}. In our model, this part is due to an interplay of the Hamiltonian dynamics and the trapping/loss dynamics, which both have their preferred basis being the site basis. The main part of the efficiency at $\lambda=35/$cm is due to incoherent Lindblad jumps, having a value of $\eta_{\rm decoherence}=83$\%. The initial state contribution is relevant only at small values of the reorganization energy.

In Fig. \ref{figContributionFMO} (center and right panel), we compare the efficiency and the coherence measure $\tilde{C}$ for the secular Redfield and the hierarchy equation of motion approach \cite{Ishizaki2009} for the Fenna-Matthews-Olson complex. The initial state is either localized at site 1 or at site 6. Four tiers of auxiliary systems were used in the computation, which already lead to a good agreement with \cite{Ishizaki2009} for the dynamics at $\lambda=35$cm$^{-1}$, $1/\gamma=50$ fs, and room temperature. In Fig. \ref{figContributionFMO} (right panel), ENAQT is observed with increasing reorganization energy also in the hierarchy approach, with the efficiency rising up to almost $\eta=100$\% at $\lambda=35/$cm.  In Fig. \ref{figContributionFMO} (center and right panel), it is observed that the normalized total coherences of the density matrix decrease with increasing reorganization energy. For the secular Redfield case, we obtain $\tilde{C}(\lambda =35$cm$^{-1})=0.0151$ for the initial site 1 and $\tilde{C}(\lambda =35$cm$^{-1})=0.0017$ for the initial site 6. For the hierarchy case, we obtain more coherence, i.e. $\tilde{C}(\lambda =35$cm$^{-1})=0.020$ for the initial site 1 and $\tilde{C}(\lambda =35$cm$^{-1})=0.0022$ for the initial site 6. In both models, coherence is more important for the rugged energy landscape of the pathway from site 1 than for the funnel-type energy landscape of the pathway from site 6.

Master equation approaches, such as the ones discussed in this section suffer from various drawbacks. Redfield theory is only applicable in the limit of weak system bath coupling and does not take into account non-equilibrium molecular reorganization effects. The hierarchy equation of motion approach assumes Gaussian fluctuations and Ohmic Drude-Lorentz spectral densities. 
The detailed atomistic structure of the protein and the chlorophylls is not taken into account in these approaches. The results thus provide a general indication of the behavior of the actual system but not a conclusive and detailed theoretical proof. In the next section, we will present a first step toward such a detailed study with our combined molecular dynamics/quantum chemistry method. The atomistic structure is included and realistic spectral densities can be obtained. We also present a straightforward method to simulate exciton dynamics beyond master equations. We thus address the second question of the microscopic origins of the long-lived quantum coherence.

\section{Molecular Dynamics Simulations}
\label{sec:md}
Among many other biologically functional components, protein complexes
are essential components of the photosynthetic system. Proteins remain
as one of the main topics of biophysical research due to their diverse
and unidentified structure-function relationship. Many biological
units are highly optimized and efficient, so that even a point
mutation of a single amino acid in conserved region often results in
the loss of the functionality. ~\cite{Tamoi2010,Jahns2002,Pesaresi1997}
Have the photosynthetic system adopted quantum mechanics to improve
its efficiency in its course of evolution? To answer this question,
careful characterization of the protein environment to the atomistic
detail is necessary to identify the microscopic origin of the long-lived
quantum coherence. As explained in the previous section, the
contribution of the quantum coherence to the energy transfer
efficiency in biological systems have been successfully carried out, yet a
more detailed description of the bath in atomic detail would be
desirable to investigate the structure-function relationship of the
protein complex and to test validity of the assumptions used in
popular models of the photosynthetic system.

The site energy of a chromophore is a complex function of the
configuration of the chromophore molecule, and the relative
orientation of the molecule to that of the embedding protein and that of  other
chromophore molecules. Factors affecting site energies have intractably
large degrees of freedom, so it is reasonable to treat those degrees
of freedom as the bath of an open quantum system. The state of the
system is assumed to be restricted to the single exciton manifold. To
construct a system-bath relationship with atomistic detail of the bath,
we start from the total Hamiltonian operator, and decomposed the
operator in such a way that the system-bath Hamiltonian is not
assumed to be any specific functional form:
\begin{equation}
  \begin{split}
  {H}_{total} &= \sum_{m} \epsilon_m (\boldsymbol{R}_{ch}, \boldsymbol{R}_{prot}) |m\rangle \langle m| + \sum_{m,n} \left\{ J_{mn} (\boldsymbol{R}_{ch}, \boldsymbol{R}_{prot}) |m\rangle \langle n| + c.c. \right\}\\
  &\quad + {T}_{ch} + {T}_{prot} + {V}_{ch}(\boldsymbol{\sigma}, \boldsymbol{R}_{ch}, \boldsymbol{R}_{prot}) + {V}_{prot}(\boldsymbol{R}_{ch}, \boldsymbol{R}_{prot}).
  \end{split}
\end{equation}
$\epsilon_m$ represents the site energy of $m$th site, $J_{mn}$ is the
coupling constant between $m$th and $n$th sites. $\boldsymbol{\sigma}$
denotes the excitonic state of chromophores, $\boldsymbol{R}_{ch}$
corresponds to the nuclear coordinates of chromophore molecules, and
$\boldsymbol{R}_{prot}$ are the nuclear coordinates of the
remaining protein and enclosing water molecules. ${T}$ and
${V}$ are the corresponding kinetic and potential energy
operators for the chromophores and proteins respectively under Born-Oppenheimer
approximation. The potential energy term for chromophores depends on
the exciton state of the systen, because dynamics of a molecule will
be governed by different Born-Oppenheimer surface when its excitonic
state changes. However, as a first approximation, we assumed that the
change of Born-Oppenheimer surfaces does not affect the bath dynamics
significantly. With this assumption, we can ignore the dependence of
the excitonic state in the $V_{ch}$ term and the system-bath
Hamiltonian only contains the one way influence from the bath to the
system:

\begin{equation}
  \begin{split}
    {H}_{total} &\approx \sum_{m} \epsilon_m (\boldsymbol{R}_{ch}, \boldsymbol{R}_{prot}) |m\rangle \langle m| + \sum_{m,n} \left\{ J_{mn} (\boldsymbol{R}_{ch}, \boldsymbol{R}_{prot}) |m\rangle \langle n| + c.c. \right\}\\
    &\quad + \sum_{m} \epsilon_m (\boldsymbol{R}_{ch}, \boldsymbol{R}_{prot}) |m\rangle \langle m| + {T}_{ch} + {T}_{prot} + {V}_{ch}(\boldsymbol{R}_{ch}, \boldsymbol{R}_{prot}) + {V}_{prot}(\boldsymbol{R}_{ch}, \boldsymbol{R}_{prot})\\ 
    &= \underbrace{ \sum_{m} \bar{\epsilon}_m |m\rangle \langle m| + \sum_{m,n} \left\{ \bar{J}_{mn} |m\rangle \langle n| + c.c. \right\} }_{{H}_S}\\&\quad + \underbrace{ \sum_{m} \left\{\epsilon_m (\boldsymbol{R}_{ch}, \boldsymbol{R}_{prot}) - \bar{\epsilon}_m \right\} |m\rangle \langle m| + \left[ \sum_{m,n} \left\{ J_{mn} (\boldsymbol{R}_{ch}, \boldsymbol{R}_{prot}) - \bar{J}_{mn} \right\} |m\rangle \langle n| + c.c. \right]  }_{{H}_{SB}}\\
    &\quad + \underbrace{ {T}_{ch} + {T}_{prot} + {V}_{ch}(\boldsymbol{R}_{ch}, \boldsymbol{R}_{prot}) + {V}_{prot}(\boldsymbol{R}_{ch}, \boldsymbol{R}_{prot}) }_{{H}_B} .
  \end{split}
\end{equation}

Based on this decomposition of the total Hamiltonian, we set up a
model of the FMO complex in atomistic detail with the AMBER force
field~\cite{Cornell1995,Ceccarelli2003} and approximate the
propagation of the entire complex by classical mechanics.
Molecular dynamics simulations were conducted at 77K and 300K with an
isothermal-isobaric (NPT) ensemble. The parameters for the system and
the system-bath Hamiltonian were calculated using quantum chemistry
methods along the trajectory from the molecular dynamics
simulation. $\epsilon_m$ was calculated using the Q-Chem quantum
chemistry package.~\cite{Shao2006} The electronic excitations were modeled
using the time-dependent density functional theory using the Tamm-Dancoff
approximation. The density functional employed was BLYP and the basis set
employed was 3-21G*. External charges from the force field were included
in the calculation as the electrostatic external potential.
The coupling terms, $J_{mn}$, were obtained from the Hamiltonian
presented in ~\cite{Adolphs2006} and considered to be constant in time.
$\bar{\epsilon}_m$ was chosen as time averaged site energy for the $m$th
site to minimize the magnitude of the system-bath Hamiltonian. In this work,
only site 1 and site 2 were considered for the exciton
dynamics. However, the methodology can be applied for the exciton
dynamic of all seven chromophores.

To obtain a closed-form equation for the reduced density matrix, we applied
mean-field approximation~\cite{May2004}; because no feedback from the system to the bath was
assumed, the state of the bath is not affected by the state of the system.
Therefore, the total density matrix, $W(t)$, can be factorized into the
reduced density matrix $\rho(t)$, and $B(t)$ which is defined only in the
Hilbert space of the bath. With additional assumption that the bath is in
thermal equilibrium, we can obtain the closed equation for the reduced density
matrix.

\begin{equation}
  \begin{split}
    \frac{\partial}{\partial t}\rho(t) &= -\frac{i}{\hbar} \left[ {H}_S, \rho(t) \right] - \frac{i}{\hbar} {\rm Tr} \left\{ \left[ {H}_{SB} , W(t) \right] \right\}\\
    &\approx -\frac{i}{\hbar} \left[ {H}_S, \rho(t) \right] - \frac{i}{\hbar} \left[ {\rm Tr} \left\{{H}_{SB} B(t)\right\} , \rho(t) \right]\\
    &\approx -\frac{i}{\hbar} \left[ {H}_S, \rho(t) \right] - \frac{i}{\hbar} \left[ {\rm Tr} \left\{{H}_{SB} B_{eq}(t)\right\} , \rho(t) \right].
  \end{split}
\end{equation}
Thermal equilibrium of the bath was ensured by the thermostat of the
molecular dynamics simulation. Thus, the reduced density matrix was
obtained by Monte Carlo integration of 4000 independent instances of
unitary quantum evolution with respect to the thermally equilibrated
bath. Each instance was propagated by integrating the Schr\"odinger
equation with the simple exponential integrator.

\begin{figure}
  \includegraphics[width=0.45\linewidth]{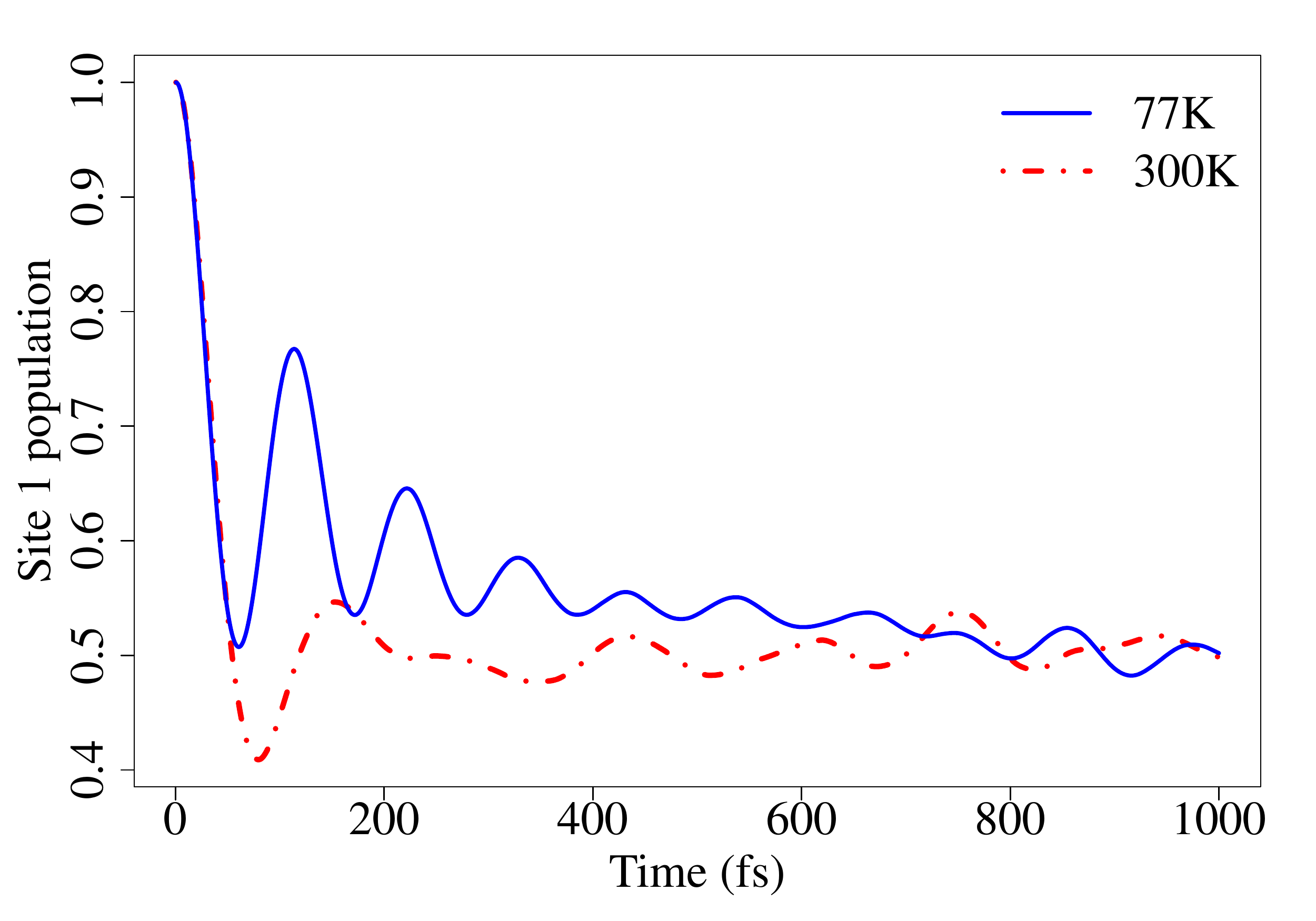}
  \includegraphics[width=0.45\linewidth]{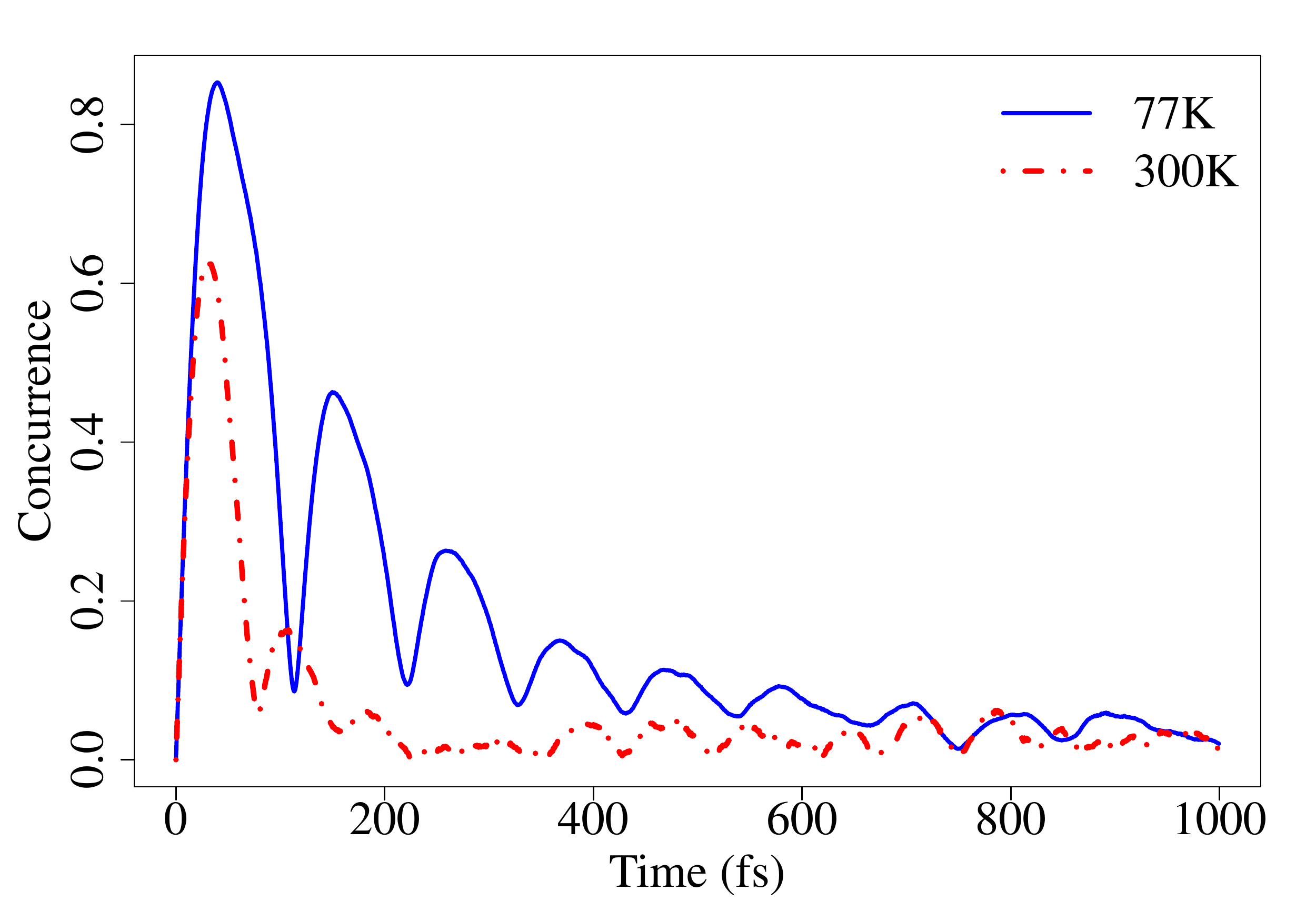}
  \caption{(Left panel) Time evolution of the exciton population at the site 1
  ($\rho_{11}$)
  based on the strongly coupled site 1 and 2 of the FMO complex at 77K and 300K.
  The initial pure state $\rho = |1\rangle \langle 1|$ was propagated
  using Monte Carlo integration of unitary evolutions, where the time-dependent site energies 
  are obtained from a combined molecular dyanmics/quantum chemistry approach.
  The asymptotic distribution does not follow a Boltzmann distribution
  because relaxation of the system to the bath is not considered.
  (Right panel) The concurrence between site 1 and 2 at 77K and 300K.
  Quantum coherence lives longer at a lower temperature.}
  \label{pop_con}
\end{figure}

Figure \ref{pop_con} shows the change of the population of the site 1, $\rho_{11}$, and the concurrence between site 1 and 2. The population is evenly distributed between the two sites because relaxation was not considered. The concurrence, $2|\rho_{12}|$, is an indicator of pairwise entanglement for the system.~\cite{Sarovar2010} Note that the coherence builds up during the first $\approx$ 100 fs , and then decreases subsequently due to the decoherence from the bath.



\begin{figure}
  \includegraphics[width=0.45\linewidth]{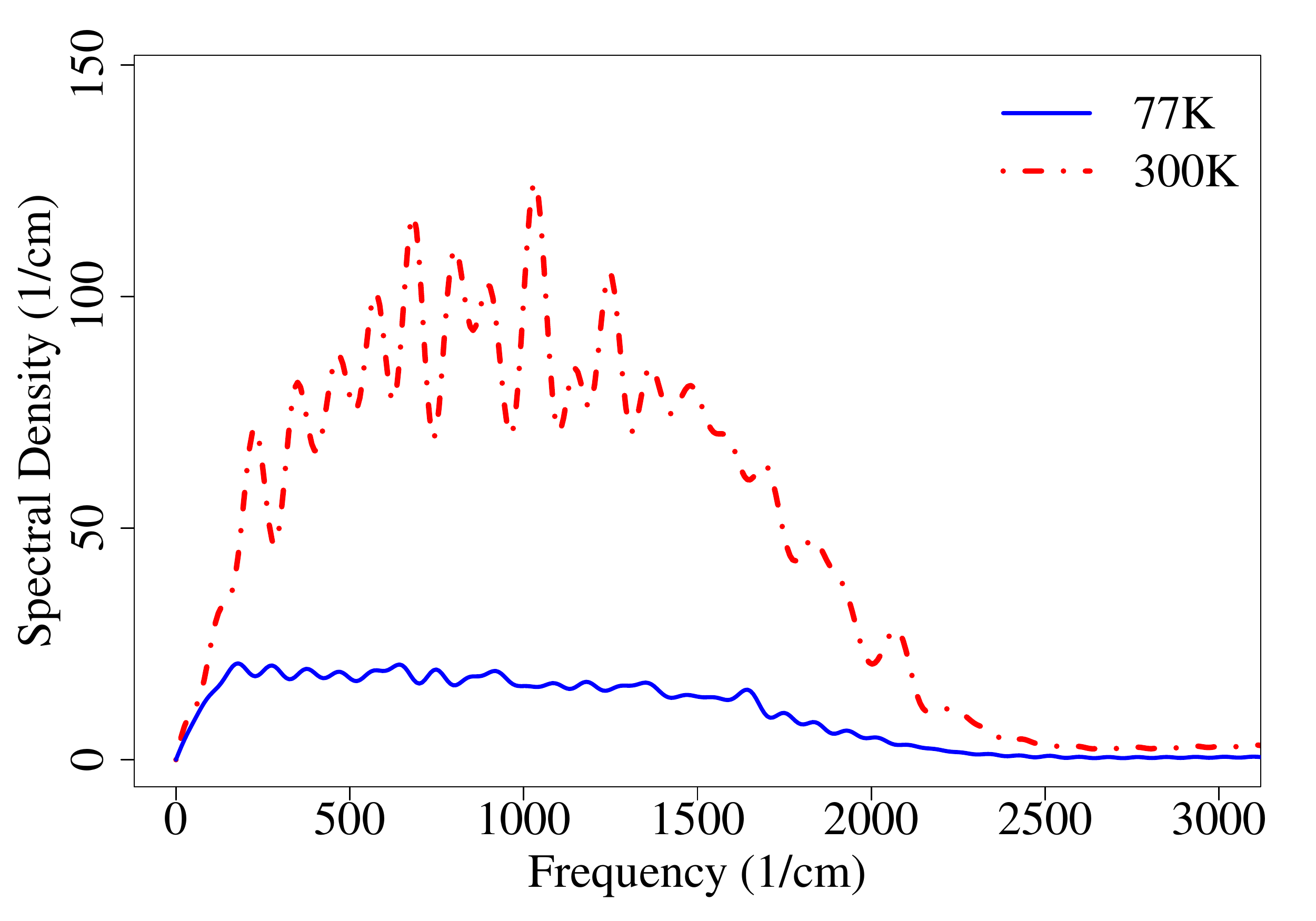}
  \includegraphics[width=0.45\linewidth]{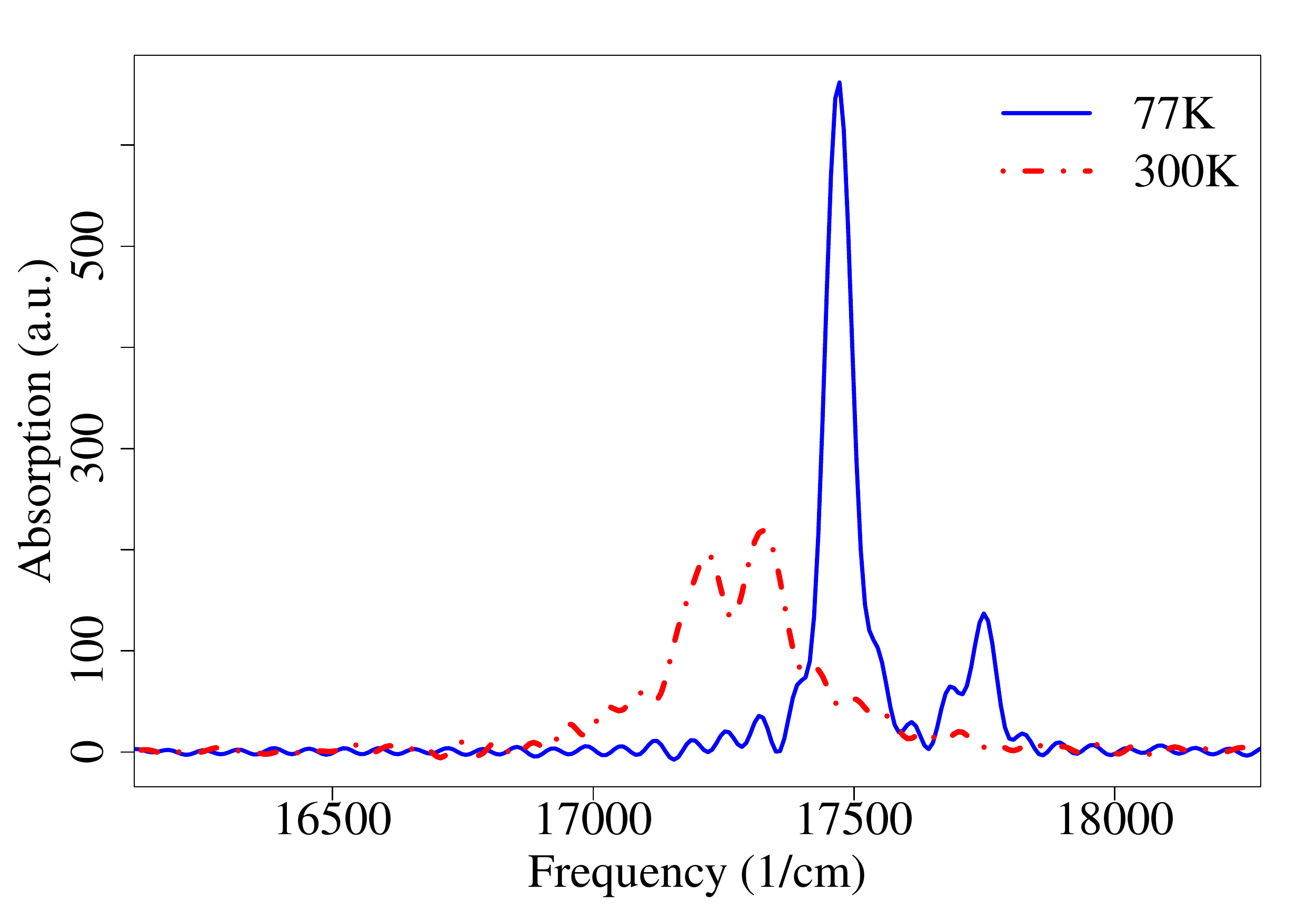}
  \caption{(Left panel) Spectral density from the autocorrelation
    function of the site 1 of the FMO complex from the molecular
    dynamics simulation at 77K and 300K. While the spectral density reflects the
    characteristic vibrational modes of the protein and the chromophore
    molecule, high-frequency modes are overpopulated due to the limitation
    of the Newtonian mechanics.
    (Right panel) Absorption spectrum of site 1 and 2 at 77K and 300K.}
  \label{Spectra}
\end{figure}


Figure \ref{Spectra} shows the spectral density of the first chromophore. Although the spectral density of the bath from molecular dynamics simulation shows characteristic frequencies related to the actual protein environment and the bacteriochloropyll molecule, high-frequency modes are overpopulated due to the limitation of the classical mechanics. There are efforts to incorporate quantum effects into the classical molecular dynamics simulation in a slightly different context,~\cite{Egorov1999,Skinner2001,Stock2009} and we are investigating the possibilities of applying these corrections.

Another simplification employed was the omission of the feedback from
exciton states. When the exciton state of a bacteriochlorophyll is
changed, the Born-Oppenheimer surface which governs the dynamics of
the chromophore molecule should be also changed. The current scheme only
propagates the protein complex on the electronically ground-state
surface. Incorporating the feedback could lead to the different
characteristics of the protein bath. There exist several schemes for mixed
quantum-classical dynamics~\cite{Tully1990,Ben-Nun2000,Wu2005} which
potentially resolve the problem at the additional computational cost of
simultaneously propagating excitons and protein bath.

Calculations are underway to carry out the full seven-site simulation
of the FMO complex at different temperatures to compare with experimental
temperature-dependent results~\cite{Panitchayangkoon2010}.

In the following final section, we will describe our quantum process
tomography scheme, which is a spectroscopic technique associated with
a computational procedure for direct extraction of the parameters related
to the quantum evolution of the system, in terms of {\sl quantum process maps}.

\section{Quantum Process Tomography}
\label{sec:qpt}

So far, we have delved into several theoretical models to characterize
quantum coherence in the entire FMO complex and in a dimer subsystem
of it. Experimentally, however, a clear characterization of this coherence
is still elusive. Signatures of long lived quantum superpositions
between excitonic states in multichromophoric systems are potentially
monitored through four wave-mixing techniques \cite{Engel2007,cheng,mukamel}.
However, a transparent description of the evolving quantum state of
the probed system is not necessarily obtained from a single realization
of such experiments. In these, a series of three weak incoming ultrashort
pulses sent from a noncollinear setup induce a macroscopic third order
polarization in the sample. The latter manifests in a time dependent
spatial grating through which a fourth pulse diffracts. From an operational
standpoint, this last pulse selects the spatial Fourier component
of the polarization which corresponds to its wavevector (heterodyne
detection), hence earning the name of four wave-mixing for this technique
(FWM) \cite{mukamel}. Extracting specific Fourier components of the
induced polarization allows for the selection of a particular set
of processes in the density matrix of the probed system, as each wavevector
is associated with a carrier frequency of the pulse. These processes
can be intuitively understood by keeping track of the dual Feynman
diagrams that account for the perturbations that the pulses induce
on the bra or ket sides of the density matrix of the probed system.
Whereas the analysis of these experiments is naturally carried out
in the density matrix formalism, an important question is whether
the density matrix itself can be imaged via these experiments, a problem
known as quantum state tomography (QST) \cite{nielsenchuang}. If
this were possible, quantum process tomography (QPT) could also be
carried out, therefore providing a complete characterization of excited
state dynamics \cite{cirac}. In a previous study, we showed that
a series of two-color heterodyned rephasing photon-echo (PE) experiments
repeated in different polarization configurations yields the necessary
information to carry out QST and QPT of the single-exciton manifold
of a coupled heterodimer \cite{YuenMohseni}. In the present article,
we adapt our previous theory to extract this information from two-dimensional
spectra, similar to those employed in current experiments.

We begin by reviewing some basic aspects of QPT. Under very general
assumptions, the evolution of an open quantum system can be described
by a linear transformation \cite{sudarshan}:

\begin{equation}
\rho_{ab}(T)=\sum_{cd}\chi_{abcd}(T)\rho_{cd}(0),\label{eq:chi tensor}\end{equation}
where $\rho_{ab}(T)$ is the element $ab$ of the reduced density
matrix $\rho$ of the system at time $T$. Equation (\ref{eq:chi tensor})
is remarkable in that $\chi(T)$ is independent of the initial state.
Knowledge of $\chi(T)$ implies a complete characterization of the
dynamics of the reduced system and, in fact, QPT can be operationally
defined as the procedure to obtain $\chi(T)$. Conceptually, it is
straightforward to recognize that, due to linearity, $\chi(T)$ can
be inverted by preparing a complete set of inputs, evolving them for
time $T$, and detecting the outputs along a complete basis. In the
context of nonlinear optical spectroscopy, this is exactly the strategy
we shall follow, with a few caveats due to experimental constraints.

To place the discussion in context, we shall be again concerned with
the subsystem composed of the excitonic dimer between sites 1 and
2 of the FMO complex. For simplicity, we ignore the rest of the sites
in this theoretical study. We only need to be concerned with four
eigenstates of this model system: The ground state $|g\rangle$, the
delocalized single-excitons $|\alpha\rangle$ and $|\beta\rangle$,
and the biexciton $|f\rangle$, which in the photosynthetic system
can be safely assumed to be the direct sum of the single-excitons
without significant interactions between them. Therefore, the biexciton
energy level is just $\omega_{f}=\omega_{\alpha}+\omega_{\beta}$.
We label the delocalized excitons so that $|\alpha\rangle$ is the
higher energy eigenstate compared to $|\beta\rangle$. Denoting the
transition energies between the $i$-th and the $j$-th states by
$\omega_{ij}=\omega_{i}-\omega_{j}$, it follows that $\omega_{\alpha g}=\omega_{f\beta}$
and $\omega_{\beta g}=\omega_{f\alpha}$ \cite{minhaengbook}. The
excitonic system is not isolated, and in fact, it interacts with a
phonon and photon bath which induces relaxation and dephasing processes
in it.

The experimental technique we consider is photon-echo (PE) spectroscopy, which is
a particular subset of FWM techniques where the wavevector of the
fourth pulse corresponds to the phase-matching condition $\boldsymbol{k}_{PE}=-\boldsymbol{k_{1}}+\boldsymbol{k_{2}}+\boldsymbol{k_{3}}$,
with $\boldsymbol{k}_{i}$ being the wavevector corresponding to the
$i$-th pulse. Here, the labeling of the pulses corresponds to the
order in which the fields interact with the sample. Typically, the
ultrashort pulses employed to study these excitonic systems possess
an optical carrier frequency, therefore allowing transitions which
are resonant with the frequency components $\pm\omega_{\beta g}$
and $\pm\omega_{\alpha g}$. In PE experiments, the first pulse centered
at $t_{1}$ creates an optical coherence beating at a frequency $\omega_{g\alpha}$
or $\omega_{g\beta}$. At $t_{2}=t_{1}+\tau$, the second pulse creates
a coherence or a population in the single exciton manifold. At $t_{3}=t_{2}+T$,
the third pulse generates another optical coherence, but this time,
beating at the frequencies opposite to the ones in the first interval,
that is, at frequencies $\omega_{\alpha g}$ or $\omega_{\beta g}$,
causing a rephasing echo of the signal. The heterodyne detection of
the nonlinear polarization signal $P_{PE}(\tau,T,t)$ occurs at time
$t_{4}=t_{3}+t$. Borrowing from NMR jargon, the intervals $(t_{1},t_{2})$,
$(t_{2},t_{3})$, and $(t_{3},t_{4})$ are traditionally refered to
as \emph{coherence}, \emph{waiting}, and \emph{echo} times, and their
durations are $\tau$, $T$, and $t$, respectively. This nomenclature
should not be taken literally. For example, in most case, coherences
do not only evolve in the coherence time, but in the waiting and echo
times. Similarly, the waiting time is often referred to as population
time, which hosts dynamics of both populations and coherences. For
a historical perspective on this vocabulary, we refer the reader to
any comprehensive NMR treatise such as \cite{goldmanbook}.

The experiment is systematically repeated for many durations for each
interval. In order to 'watch' single-exciton dynamics, it is convenient
to isolate the changes on the signal due to the waiting time $T$.
This exercise is accomplished by performing a double Fourier transform
of the signal along the $\tau$ and $t$ axes, which yields a 2D spectra
that evolves in $T$ \cite{ginsberg,jonas,minhaeng}:

\begin{equation}
S(\omega_{\tau},T,\omega_{T})=\int_{0}^{\infty}d\tau\int_{0}^{\infty}dtP_{PE}(\tau,T,t)e^{-i\omega_{\tau}\tau+i\omega_{T}T}\label{eq:ft of polarization}\end{equation}

In order to map a PE experiment to a QPT, we identify the coherence
interval as the preparation step and the echo interval as the detection
step. This assumption implies that the optical coherence intervals
have well characterized dynamics. This hypothesis is reasonable due
to a separation of timescales where optical coherences will presumably
decay exponentially due to pure dephasing and not due to intricate
phonon-induced processes. Therefore, the 2D spectrum consists of four
Lorentzian peaks centered about $(\omega_{\tau},\omega_{t})=(\omega_{\alpha g},\omega_{\alpha g}),(\omega_{\alpha g},\omega_{\beta g}),(\omega_{\beta g},\omega_{\alpha g}),(\omega_{\beta g},\omega_{\beta g})$.
In this discussion, we shall ignore inhomogeneous broadening, noting
that it can always be accounted for as a convolution of the signal
with the distrubtion of inhomogeneity. The width of these Lorentzians
can be directly related to the dephasing rates of the optical coherences.
Loosely speaking, a particular value on the $\omega_{\tau}$ axis
of the spectrum indicates a specific type of state preparation, whereas
the $\omega_{t}$ axis is related to a particular detection. More
precisely, a peak in the 2D spectrum displays the correlations between
the frequency beats from the coherence and echo intervals. A crucial
realization is that the amplitude of these peaks can be written as
a linear combination of elements of the time evolving excitonic density
matrix stemming from different initial states, that is, of elements
of $\chi(T)$ itself \cite{YuenMohseni}:

\begin{eqnarray}
\tilde{S}(\omega_{\alpha g},T,\omega_{\alpha g}) & = & -C_{\omega_{1}}^{\alpha}C_{\omega_{2}}^{\alpha}(\boldsymbol{\mu}_{\alpha g}\cdot\boldsymbol{e}_{1})(\boldsymbol{\mu}_{\alpha g}\cdot\boldsymbol{e}_{2})\nonumber \\
 &  & \times\{C_{\omega_{3}}^{\alpha}[(\boldsymbol{\mu}_{\alpha g}\cdot\boldsymbol{e}_{3})(\boldsymbol{\mu}_{\alpha g}\cdot\boldsymbol{e}_{4})(\chi_{gg\alpha\alpha}(T)-1-\chi_{\alpha\alpha\alpha\alpha}(T))\nonumber \\
 &  & +(\boldsymbol{\mu}_{f\beta}\cdot\boldsymbol{e}_{3})(\boldsymbol{\mu}_{f\beta}\cdot\boldsymbol{e}_{4})\chi_{\beta\beta\alpha\alpha}(T)]\nonumber \\
 &  & +C_{\omega_{3}}^{\beta}[(\boldsymbol{\mu}_{f\alpha}\cdot\boldsymbol{e}_{3})(\boldsymbol{\mu}_{f\beta}\cdot\boldsymbol{e}_{4})-(\boldsymbol{\mu}_{\beta g}\cdot\boldsymbol{e}_{3})(\boldsymbol{\mu}_{\alpha g}\cdot\boldsymbol{e}_{4}))\chi_{\alpha\beta\alpha\alpha}(T)]\}\nonumber \\
 &  & -C_{\omega_{1}}^{\alpha}C_{\omega_{2}}^{\beta}(\boldsymbol{\mu}_{\alpha g}\cdot\boldsymbol{e}_{1})(\boldsymbol{\mu}_{\beta g}\cdot\boldsymbol{e}_{2})\nonumber \\
 &  & \times\{C_{\omega_{3}}^{\alpha}[(\boldsymbol{\mu}_{\alpha g}\cdot\boldsymbol{e}_{3})(\boldsymbol{\mu}_{\alpha g}\cdot\boldsymbol{e}_{4})(\chi_{gg\beta\alpha}(T)-\chi_{\alpha\alpha\beta\alpha}(T))\nonumber \\
 &  & +(\boldsymbol{\mu}_{f\beta}\cdot\boldsymbol{e}_{3})(\boldsymbol{\mu}_{f\beta}\cdot\boldsymbol{e}_{4})\chi_{\beta\beta\beta\alpha}(T)]\nonumber \\
 &  & +C_{\omega_{3}}^{\beta}[((\boldsymbol{\mu}_{f\alpha}\cdot\boldsymbol{e}_{3})(\boldsymbol{\mu}_{f\beta}\cdot\boldsymbol{e}_{4})-(\boldsymbol{\mu}_{\beta g}\cdot\boldsymbol{e}_{3})(\boldsymbol{\mu}_{\alpha g}\cdot\boldsymbol{e}_{4}))\chi_{\alpha\beta\beta\alpha}(T)]\},\label{eq:peakamplitudealphaalpha}\end{eqnarray}

\begin{eqnarray}
\tilde{S}(\omega_{\alpha g},T,\omega_{\beta g}) & = & -C_{\omega_{1}}^{\alpha}C_{\omega_{2}}^{\alpha}(\boldsymbol{\mu}_{\alpha g}\cdot\boldsymbol{e}_{1})(\boldsymbol{\mu}_{\alpha g}\cdot\boldsymbol{e}_{2})\nonumber \\
 &  & \times\{C_{\omega_{3}}^{\beta}[(\boldsymbol{\mu}_{\beta g}\cdot\boldsymbol{e}_{3})(\boldsymbol{\mu}_{\beta g}\cdot\boldsymbol{e}_{4})(\chi_{gg\alpha\alpha}(T)-1-\chi_{\beta\beta\alpha\alpha}(T))\nonumber \\
 &  & +(\boldsymbol{\mu}_{f\alpha}\cdot\boldsymbol{e}_{3})(\boldsymbol{\mu}_{f\alpha}\cdot\boldsymbol{e}_{4})\chi_{\alpha\alpha\alpha\alpha}(T)]\nonumber \\
 &  & +C_{\omega_{3}}^{\alpha}[((\boldsymbol{\mu}_{f\beta}\cdot\boldsymbol{e}_{3})(\boldsymbol{\mu}_{f\alpha}\cdot\boldsymbol{e}_{4})-(\boldsymbol{\mu}_{\alpha g}\cdot\boldsymbol{e}_{3})(\boldsymbol{\mu}_{\beta g}\cdot\boldsymbol{e}_{4}))\chi_{\beta\alpha\alpha\alpha}(T)]\}\nonumber \\
 &  & -C_{\omega_{1}}^{\alpha}C_{\omega_{2}}^{\beta}(\boldsymbol{\mu}_{\alpha g}\cdot\boldsymbol{e}_{1})(\boldsymbol{\mu}_{\beta g}\cdot\boldsymbol{e}_{2})\nonumber \\
 &  & \times\{C_{\omega_{3}}^{\beta}[(\boldsymbol{\mu}_{\beta g}\cdot\boldsymbol{e}_{3})(\boldsymbol{\mu}_{\beta g}\cdot\boldsymbol{e}_{4})(\chi_{gg\beta\alpha}(T)-\chi_{\beta\beta\beta\alpha}(T))\nonumber \\
 &  & +(\boldsymbol{\mu}_{f\alpha}\cdot\boldsymbol{e}_{3})(\boldsymbol{\mu}_{f\alpha}\cdot\boldsymbol{e}_{4})\chi_{\alpha\alpha\beta\alpha}(T)]\nonumber \\
 &  & +C_{\omega_{3}}^{\alpha}[((\boldsymbol{\mu}_{f\beta}\cdot\boldsymbol{e}_{3})(\boldsymbol{\mu}_{f\alpha}\cdot\boldsymbol{e}_{4})-(\boldsymbol{\mu}_{\alpha g}\cdot\boldsymbol{e}_{3})(\boldsymbol{\mu}_{\beta g}\cdot\boldsymbol{e}_{4}))\chi_{\beta\alpha\beta\alpha}(T)]\},\label{eq:peakamplitudealphabeta}\end{eqnarray}

\begin{eqnarray}
\tilde{S}(\omega_{\beta g},T,\omega_{\alpha g}) & = & -C_{\omega_{1}}^{\beta}C_{\omega_{2}}^{\beta}(\boldsymbol{\mu}_{\beta g}\cdot\boldsymbol{e}_{1})(\boldsymbol{\mu}_{\beta g}\cdot\boldsymbol{e}_{2})\nonumber \\
 &  & \times\{C_{\omega_{3}}^{\alpha}[(\boldsymbol{\mu}_{\alpha g}\cdot\boldsymbol{e}_{3})(\boldsymbol{\mu}_{\alpha g}\cdot\boldsymbol{e}_{4})(\chi_{gg\beta\beta}(T)-1-\chi_{\alpha\alpha\beta\beta}(T))\nonumber \\
 &  & +(\boldsymbol{\mu}_{f\beta}\cdot\boldsymbol{e}_{3})(\boldsymbol{\mu}_{f\beta}\cdot\boldsymbol{e}_{4})\chi_{\beta\beta\beta\beta}(T)]\nonumber \\
 &  & +C_{\omega_{3}}^{\beta}[(\boldsymbol{\mu}_{f\alpha}\cdot\boldsymbol{e}_{3})(\boldsymbol{\mu}_{f\beta}\cdot\boldsymbol{e}_{4})-(\boldsymbol{\mu}_{\beta g}\cdot\boldsymbol{e}_{3})(\boldsymbol{\mu}_{\alpha g}\cdot\boldsymbol{e}_{4}))\chi_{\alpha\beta\beta\beta}(T)]\}\nonumber \\
 &  & -C_{\omega_{1}}^{\beta}C_{\omega_{2}}^{\alpha}(\boldsymbol{\mu}_{\beta g}\cdot\boldsymbol{e}_{1})(\boldsymbol{\mu}_{\alpha g}\cdot\boldsymbol{e}_{2})\nonumber \\
 &  & \times\{C_{\omega_{3}}^{\alpha}[(\boldsymbol{\mu}_{\alpha g}\cdot\boldsymbol{e}_{3})(\boldsymbol{\mu}_{\alpha g}\cdot\boldsymbol{e}_{4})(\chi_{gg\alpha\beta}(T)-\chi_{\alpha\alpha\alpha\beta}(T))\nonumber \\
 &  & +(\boldsymbol{\mu}_{f\beta}\cdot\boldsymbol{e}_{3})(\boldsymbol{\mu}_{f\beta}\cdot\boldsymbol{e}_{4})\chi_{\beta\beta\alpha\beta}(T)]\nonumber \\
 &  & +C_{\omega_{3}}^{\beta}[((\boldsymbol{\mu}_{f\alpha}\cdot\boldsymbol{e}_{3})(\boldsymbol{\mu}_{f\beta}\cdot\boldsymbol{e}_{4})-(\boldsymbol{\mu}_{\beta g}\cdot\boldsymbol{e}_{3})(\boldsymbol{\mu}_{\alpha g}\cdot\boldsymbol{e}_{4}))\chi_{\alpha\beta\alpha\beta}(T)]\},\label{eq:peakamplitudebetaalpha}\end{eqnarray}

\begin{eqnarray}
\tilde{S}(\omega_{\beta g},T,\omega_{\beta g}) & = & -C_{\omega_{1}}^{\beta}C_{\omega_{2}}^{\beta}(\boldsymbol{\mu}_{\beta g}\cdot\boldsymbol{e}_{1})(\boldsymbol{\mu}_{\beta g}\cdot\boldsymbol{e}_{2})\nonumber \\
 &  & \times\{C_{\omega_{3}}^{\beta}[(\boldsymbol{\mu}_{\beta g}\cdot\boldsymbol{e}_{3})(\boldsymbol{\mu}_{\beta g}\cdot\boldsymbol{e}_{4})(\chi_{gg\beta\beta}(T)-1-\chi_{\beta\beta\beta\beta}(T))\nonumber \\
 &  & +(\boldsymbol{\mu}_{f\alpha}\cdot\boldsymbol{e}_{3})(\boldsymbol{\mu}_{f\alpha}\cdot\boldsymbol{e}_{4})\chi_{\alpha\alpha\beta\beta}(T)]\nonumber \\
 &  & +C_{\omega_{3}}^{\alpha}[((\boldsymbol{\mu}_{f\beta}\cdot\boldsymbol{e}_{3})(\boldsymbol{\mu}_{f\alpha}\cdot\boldsymbol{e}_{4})-(\boldsymbol{\mu}_{\alpha g}\cdot\boldsymbol{e}_{3})(\boldsymbol{\mu}_{\beta g}\cdot\boldsymbol{e}_{4}))\chi_{\beta\alpha\beta\beta}(T)]\}\nonumber \\
 &  & -C_{\omega_{1}}^{\beta}C_{\omega_{2}}^{\alpha}(\boldsymbol{\mu}_{\beta g}\cdot\boldsymbol{e}_{1})(\boldsymbol{\mu}_{\alpha g}\cdot\boldsymbol{e}_{2})\nonumber \\
 &  & \times\{C_{\omega_{3}}^{\beta}[(\boldsymbol{\mu}_{\beta g}\cdot\boldsymbol{e}_{3})(\boldsymbol{\mu}_{\beta g}\cdot\boldsymbol{e}_{4})(\chi_{gg\alpha\beta}(T)-\chi_{\beta\beta\alpha\beta}(T))\nonumber \\
 &  & +(\boldsymbol{\mu}_{f\alpha}\cdot\boldsymbol{e}_{3})(\boldsymbol{\mu}_{f\alpha}\cdot\boldsymbol{e}_{4})\chi_{\alpha\alpha\alpha\beta}(T)]\nonumber \\
 &  & +C_{\omega_{3}}^{\alpha}[((\boldsymbol{\mu}_{f\beta}\cdot\boldsymbol{e}_{3})(\boldsymbol{\mu}_{f\alpha}\cdot\boldsymbol{e}_{4})-(\boldsymbol{\mu}_{\alpha g}\cdot\boldsymbol{e}_{3})(\boldsymbol{\mu}_{\beta g}\cdot\boldsymbol{e}_{4}))\chi_{\beta\alpha\alpha\beta}(T)]\}.\label{eq:peakamplitudebetabeta}\end{eqnarray}
Here, the expressions have been obtained using the rotating-wave approximation,
as well as the assumption of no overlap between pulses. $\boldsymbol{\mu}_{pq}=\boldsymbol{\mu}_{qp}$
is the transition dipole moment between states $p,q\in\{g,\alpha,\beta,f\}$. We have
rescaled the spectra amplitudes to eliminate the details of the lineshape
by multiplying them by the dephasing rates of the optical coherences
in the coherence and echo intervals,

\begin{equation}
\tilde{S}(\omega_{pg},T,\omega_{qg})=\Gamma_{gp}\Gamma_{qg}S(\omega_{pg},T,\omega_{qg}).\label{eq:rescaling}\end{equation}
The coefficient $C_{\omega_{i}}^{p}$ is the amplitude of the $i$-th
pulse at the frequency $\omega_{pg}$,

\begin{equation}
C_{\omega_{i}}^{p}=-\frac{\Lambda}{i}\sqrt{2\pi\sigma^{2}}e^{-\sigma^{2}(\omega_{pg}-\omega_{i})^{2}/2},\label{eq:coefficient}\end{equation}
with $\Lambda$ being the strength of the pulse and $\sigma$ the
width of the Gaussian pulse in time domain. Also, $\boldsymbol{e}_{i}$
is the polarization of the $i$-th pulse. Both $C_{\omega_{i}}^{p}$
and $\boldsymbol{e}_{i}$ are experimentally tunable parameters for
the pulses.

Whereas Equations (14) and (15) presented in \cite{YuenMohseni} correspond
to a single value of $\tau$ and $t$, Equations (\ref{eq:peakamplitudealphaalpha}),
(\ref{eq:peakamplitudealphabeta}), (\ref{eq:peakamplitudebetaalpha}),
and (\ref{eq:peakamplitudebetabeta}) stem from Fourier transform
of data collected at many $\tau$ and $t$ times. Therefore, in principle,
a 2D spectrum provides a more robust source of information from which
to invert $\chi(T)$ than in the suggested 1D experiment. The displayed
equations, albeit lengthy, are easy to interpret. For instance, consider
the term which is proportional to $\chi_{\alpha\beta\alpha\alpha}(T)$
in Equation (\ref{eq:peakamplitudealphaalpha}), which stems from
the Feynman diagram depicted in Fig. 5. As expected, it consists of
a waiting time where the initially prepared population $|\alpha\rangle\langle\alpha|$
is transferred to the coherence $|\alpha\rangle\langle\beta|$. This
waiting time is escorted by a coherence $|g\rangle\langle\alpha|$
oscillating as $e^{(-i\omega_{g\alpha}-\Gamma_{g\alpha})\tau}$ which
evolves during the coherence time and another set of coherences $|f\rangle\langle\beta|$
and $|\alpha\rangle\langle g|$ which evolve during the echo time
as $e^{(-i\omega_{f\beta}-\Gamma_{f\beta})t}=e^{(-i\omega_{\alpha g}-\Gamma_{\alpha g})t}$.
These two intervals correspond to the diagonal peak located at $(\omega_{\alpha g},\omega_{\alpha g})$.
Other processes that exhibit oscillations at those two respective
frequencies appear as additional terms in the equation corresponding
to that particular peak.

\begin{center}
\begin{figure}
\begin{centering}
\includegraphics[scale=0.5]{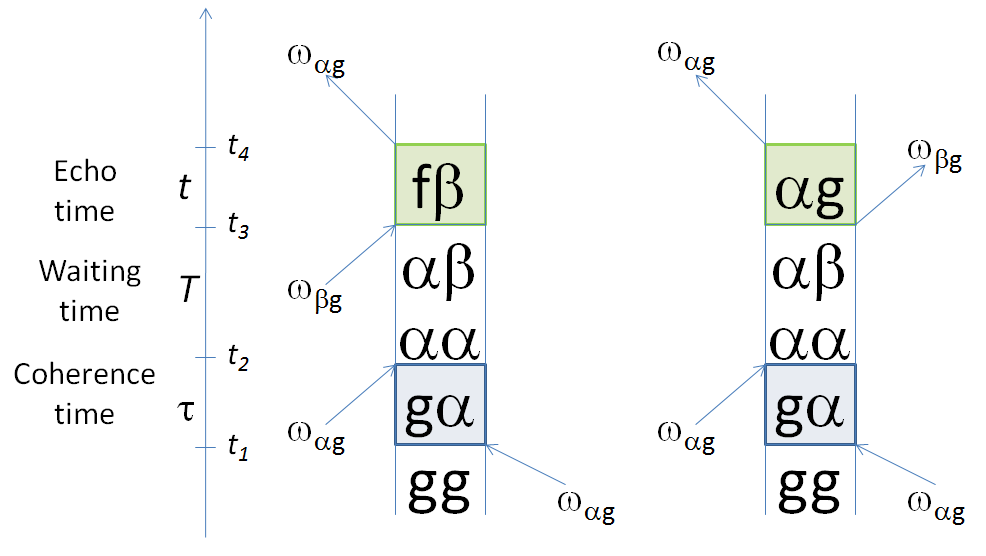}
\par\end{centering}

\caption{Dual Feynman diagrams that account for the population to coherence
transfer terms $\chi_{\alpha\beta\alpha\alpha}(T)$ in quantum process
tomography.}

\end{figure}

\par\end{center}

In Ref. \cite{YuenMohseni}, we showed that there are sixteen real valued
parameters of $\chi(T)$ which need to be determined at every value
of $T$ in order to carry out QPT of the single exciton manifold of
a heterodimer. For an illustration, we shall describe how to obtain
the elements $\chi_{ij\alpha\alpha}(T)$. These quantities are shown
in Fig. 6 and have been computed using the Ishizaki-Fleming model,
with a bath correlation of 150 $fs$ \cite{Ishizaki2009}.
They display rich and nontrivial phonon-induced behavior, such as
the spontaneous generation of coherence from a population in an eigenstate
of the excitonic Hamiltonian, and therefore, is a very good example
of how QPT provides access to this nontrivial information via the
repetition of a series of 2D PE experiments. For this particular set
of $\chi(T)$ elements, we shall exploit the waveform of the pulses
but not their polarizations, and for simplicitly we will assume the
polarization configuration $xxxx$ for each of the pulses including
the heterodyning.

\begin{center}
\begin{figure}
\begin{centering}
\includegraphics[scale=1.0]{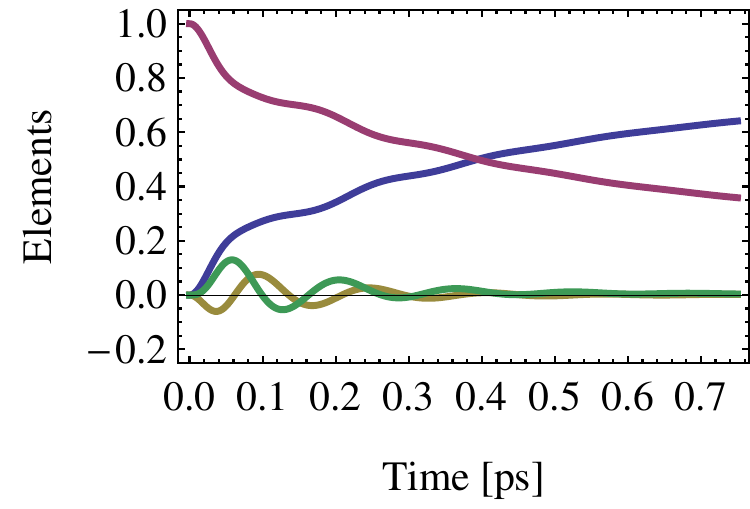}
\par\end{centering}

\caption{Transfer of population in eigenstate $|\alpha\rangle\langle\alpha|$ to other populations and coherences in the eigenbasis of the single exciton Hamiltonian. 
The hierarchy equation of motion approach is used for a dimer system based on the parameters of the site 1 and site 2 subsystem of the Fenna-Matthews-Olson complex. Population in $|\alpha\rangle\langle\alpha|$ decreases
($\chi_{\alpha\alpha\alpha\alpha}(T)$, purple) and is transferred
to $|\beta\rangle\langle\beta|$ ($\chi_{\beta\beta\alpha\alpha}(T)$,
blue). Emergence of coherence from the initial population
occurs in this model ($\Re\{\chi_{\alpha\beta\alpha\alpha}(T)\}$,
yellow and $\Im\{\chi_{\alpha\alpha\alpha\alpha}(T)\}$, green).}

\end{figure}

\par\end{center}

Consider the possibility of using pulses with carrier frequencies
centered about $\omega_{\alpha g}$ and $\omega_{\beta g}$ respectively,
and such that their bandwidth is narrow enough that the pulse centered
about $\omega_{\alpha g}$ has negligible component at $\omega_{\beta g}$
and vice versa. Then, we can carry out an experiment such that $\frac{|C_{\omega_{1}}^{\alpha}|}{|C_{\omega_{1}}^{\beta}|},\frac{|C_{\omega_{2}}^{\alpha}|}{|C_{\omega_{2}}^{\beta}|},\frac{|C_{\omega_{3}}^{\beta}|}{|C_{\omega_{3}}^{\alpha}|}\gg1$
(experiment 1) for all $i$ and notice that the diagonal peak at $(\omega_{\alpha g},\omega_{\alpha g})$
reduces to:

\begin{eqnarray}
\langle\tilde{S}(\omega_{\alpha},T,\omega_{\alpha})\rangle_{xxxx}=-C_{\omega_{1}}^{\alpha}C_{\omega_{2}}^{\alpha}C_{\omega_{3}}^{\beta} & \langle(\boldsymbol{\mu}_{\alpha g}\cdot\boldsymbol{e}_{1})(\boldsymbol{\mu}_{\alpha g}\cdot\boldsymbol{e}_{2})[(\boldsymbol{\mu}_{f\alpha}\cdot\boldsymbol{e}_{3})(\boldsymbol{\mu}_{f\beta}\cdot\boldsymbol{e}_{4})\\ \nonumber & -(\boldsymbol{\mu}_{\beta g}\cdot\boldsymbol{e}_{3})(\boldsymbol{\mu}_{\alpha g}\cdot\boldsymbol{e}_{4}))]\rangle_{xxxx} \chi_{\alpha\beta\alpha\alpha}(T)\label{eq:directmonitor}\end{eqnarray}
which implies that its evolution with respect to $T$ directly monitors
the transfer of the population prepared at $|\alpha\rangle\langle\alpha|$
to the coherence at $|\alpha\rangle\langle\beta|$. Here, $\langle\cdot\rangle_{xxxx}$
denotes an isotropic average of the experiments performed with the
$xxxx$ polarization configuration. $\chi_{\alpha\beta\alpha\alpha}(T)$
can be directly obtained if information of the dipole moments is known
in advance. As can be checked easily, $\chi_{\alpha\beta\alpha\alpha}(T)=(\chi_{\beta\alpha\alpha\alpha}(T))^{*}$
can, in principle, be also obtained directly from an experiment where
$\frac{|C_{\omega_{i}}^{\alpha}|}{|C_{\omega_{i}}^{\beta}|}\gg1$
for all $i$ (experiment 2) and monitoring $\langle\tilde{S}(\omega_{\alpha},T,\omega_{\beta})\rangle_{xxxx}$.
Redundant measurements can be used as ways of effectively constraining
the QPT.

Similarly, the transfer from $|\alpha\rangle\langle\alpha|$ to other
populations can be extracted by monitoring $\langle\tilde{S}(\omega_{\alpha},T,\omega_{\alpha})\rangle_{xxxx}$
in experiment 2 and $\langle\tilde{S}(\omega_{\alpha},T,\omega_{\beta})\rangle_{xxxx}$
in experiment 1. These two linearly independent conditions are enough
to extract $\chi_{gg\alpha\alpha}(T)$, $\chi_{\alpha\alpha\alpha\alpha}(T)$,
and $\chi_{\beta\beta\alpha\alpha}(T)$, since there is a third independent
condition based on trace preservation which reads $\chi_{gg\alpha\alpha}(T)+\chi_{\alpha\alpha\alpha\alpha}(T)+\chi_{\beta\beta\alpha\alpha}(T)=1$.

It is now important to verify whether the suggested experiments are
feasible. In order to ensure conditions of the form $\frac{|C_{\omega_{i}}^{\alpha}|}{|C_{\omega_{i}}^{\beta}|}\gg1$,
we need $\sigma\sim\frac{3}{\omega_{\alpha g}-\omega_{\beta g}}\sim470$ fs,
that is, the pulse needs to be long enough to guarantee the narrow
band condition. This requirement, although attractive from a pedagogical
standpoint since it yields block diagonal sets of linear equations,
is a nuisance from a practical perspective, as decoherence mechanisms
might be in the same timescale and might not be 'seen' with such long
pulses. However, the only essential requirement is a toolbox of two
different waveforms for the pulses. A more sensible choice is a set
of pulses centered about $\omega_{\alpha g}$ and $\omega_{\beta g}$
respectively, but having $\sigma\sim100$  fs. By carrying out 8
experiments alternating the two waveforms in each of the three pulses,
each of the terms in Equations (\ref{eq:peakamplitudealphaalpha}),
(\ref{eq:peakamplitudealphabeta}), (\ref{eq:peakamplitudebetaalpha}),
and (\ref{eq:peakamplitudebetabeta}) which are proportional to $C_{\omega_{1}}^{i}C_{\omega_{2}}^{j}C_{\omega_{3}}^{k}$
for $i,j,k\in\{\alpha,\beta\}$ may be inverted to yield the block
diagonal set of equations discussed above.

In summary, we have presented three different tools for unraveling the
role of quantum coherence in biological systems: a) techniques for obtaining the contribution of quantum coherences to
biological processes; b) a microscopic simulation approach to explore
the dynamics of these systems by direct simulation; and finally c) a
new theoretical proposal for an experimental procedure that provides
detailed information about the quantum procesess associated with
energy transfer in the ultrafast regime. We believe that ultimately, a
combination of these three techniques and tools from other
groups will be collectively required to make definitive conclusions about
the role of quantum coherence in photosynthetic complexes.

We are thankful for financial support from the Department of Energy Frontier
Research Center for Excitonics, the Army Research Office and the Sloan and
Camille and Henry Dreyfus foundations.






\bibliographystyle{apsrev4-1}
\bibliography{PatricksPapers,SangwoosPapers,JoelsPapers}

\begin{thebibliography}{10}
\expandafter\ifx\csname url\endcsname\relax
  \def\url#1{\texttt{#1}}\fi
\expandafter\ifx\csname urlprefix\endcsname\relax\def\urlprefix{URL }\fi
\expandafter\ifx\csname href\endcsname\relax
  \def\href#1#2{#2} \def\path#1{#1}\fi

\bibitem{Blankenship2002}
R.E. Blankenship, {Molecular Mechanisms of Photosynthesis}, 1st Edition,
  Wiley-Blackwell, 2002.

\bibitem{Cheng2009}
Y.-C. Cheng, G.R. Fleming, 
  Ann. Rev. Phys. Chem. 60 (2009) 241.

\bibitem{Engel2007}
G.S. Engel, T.R. Calhoun, E.L. Read, T.-K. Ahn, T.~Mancal, Y.-C. Cheng,
  R.E. Blankenship, G.R. Fleming, 
Nature 446~(7137) (2007) 782.

\bibitem{Lee2007}
H.~Lee, Y.-C. Cheng, G.R. Fleming, 
Science (New York, N.Y.) 316~(5830) (2007) 1462.

\bibitem{Panitchayangkoon2010}
G.~Panitchayangkoon, D.~Hayes, K.A. Fransted, J.R. Caram, E.~Harel, J.~Wen,
  R.E. Blankenship, G.S. Engel, 
Proc. Natl. Acad.
  Sci. USA (2010) 1.

\bibitem{Collini2010}
E.~Collini, C.Y. Wong, K.E. Wilk, P.M.G. Curmi, P.~Brumer, G.D. Scholes,
 Nature 463~(7281) (2010) 644.

\bibitem{Rebentrost2009b}
P.~Rebentrost, M.~Mohseni, I.~Kassal, S.~Lloyd, A.~Aspuru-Guzik,
New Journal of Physics 11~(3)
  (2009) 033003.

\bibitem{Plenio2008}
M.B. Plenio, S.F. Huelga, 
New J. Phys. 10~(11) (2008) 113019.

\bibitem{Mohseni2008}
M.~Mohseni, P.~Rebentrost, S.~Lloyd, A.~Aspuru-Guzik, 
J. Chem. Phys 129~(17)
  (2008) 174106.

\bibitem{Ishizaki2009}
A.~Ishizaki, G.R. Fleming, 
Proc. Natl. Acad. Sci.
  USA 106~(41) (2009) 17255.

\bibitem{Jang2008}
S.~Jang, Y.-C. Cheng, D.R. Reichman, J.D. Eaves, 
J. Chem. Phys. 129~(10) (2008) 101104.

\bibitem{Wu2010}
J.~Wu, F.~Liu, Y.~Shen, J.~Cao, R.J. Silbey,
unpublished.
\urlprefix\url{http://arxiv.org/abs/1008.2236}.

\bibitem{Sarovar2010}
M.~Sarovar, A.~Ishizaki, G.R. Fleming, K.B. Whaley, 
Nature Physics 6~(6) (2010)
  462.

\bibitem{Caruso2010a}
F.~Caruso, A.~Chin, A.~Datta, S.~Huelga, M.~Plenio, 
Phys. Rev. A
  81~(6) (2010) 1.

\bibitem{Fassioli2010}
F.~Fassioli, A.~Olaya-Castro, 
New J. Phys. 12 (2010) 085006.

\bibitem{Cai2010}
J.~Cai, G.G. Guerreschi, H.J. Briegel, 
Phys. Rev. Lett. 104~(22) (2010) 1.

\bibitem{Rebentrost2009a}
P.~Rebentrost, M.~Mohseni, A.~Aspuru-Guzik, 
J. Phys. Chem.
  B 113~(29) (2009) 9942.

\bibitem{Olaya-Castro2008}
A.~Olaya-Castro, C.~Lee, F.~Olsen, N.~Johnson, 
Phys. Rev. B 78~(8)
  (2008) 7.

\bibitem{Shi2009}
Q.~Shi, L.~Chen, G.~Nan, R.-X. Xu, Y.~Yan, 
J. Chem. Phys. 130~(8)
  (2009) 084105.

\bibitem{Zhu2010}
J.~Zhu, S.~Kais, P.~Rebentrost, A.~Aspuru-Guzik, 
submitted.

\bibitem{Piilo2008}
J.~Piilo, S.~Maniscalco, K.~H\"{a}rk\"{o}nen, K.-A. Suominen, 
Phys. Rev. Lett. 100~(18) (2008) 1.

\bibitem{Rebentrost2009}
P.~Rebentrost, R.~Chakraborty, A.~Aspuru-Guzik, 
J. Chem. Phys. 131~(18) (2009) 184102.

\bibitem{Adolphs2006}
J.~Adolphs, T.~Renger, 
Biophys. J. 91~(8) (2006) 2778.

\bibitem{Tamoi2010}
M.~Tamoi, T.~Tabuchi, M.~Demuratani, K.~Otori, N.~Tanabe, T.~Maruta,
  S.~Shigeoka, 
J. Bio. Chem. 285~(20) (2010)
  15399.

\bibitem{Jahns2002}
P.~Jahns, M.~Graf, Y.~Munekage, T.~Shikanai, 
FEBS letters 519~(1-3)
  (2002) 99.

\bibitem{Pesaresi1997}
P.~Pesaresi, D.~Sandon\`{a}, E.~Giuffra, R.~Bassi, 
FEBS Letters 402~(2-3) (1997) 151.

\bibitem{Cornell1995}
W.D. Cornell, P.~Cieplak, C.I. Bayly, I.R. Gould, K.M. Merz, D.M.
  Ferguson, D.C. Spellmeyer, T.~Fox, J.W. Caldwell, P.A. Kollman, 
J. Am. Chem. Soc. 117~(19) (1995) 5179.

\bibitem{Ceccarelli2003}
M.~Ceccarelli, P.~Procacci, M.~Marchi, 
J. Comp. Chem. 24~(2) (2003)
  129.

\bibitem{Shao2006}
Y.~Shao, L.F. Molnar, Y.~Jung, J.~Kussmann, C.~Ochsenfeld, S.T. Brown,
  A.T.B. Gilbert, L.V. Slipchenko, S.V. Levchenko, D.P. O'Neill, R.A.
  DiStasio, R.C. Lochan, T.~Wang, G.J.O. Beran, N.A. Besley, J.M. Herbert,
  C.Y. Lin, T.~{Van Voorhis}, S.H. Chien, A.~Sodt, R.P. Steele, V.A.
  Rassolov, P.E. Maslen, P.P. Korambath, R.D. Adamson, B.~Austin, J.~Baker,
  E.F.C. Byrd, H.~Dachsel, R.J. Doerksen, A.~Dreuw, B.D. Dunietz, A.D.
  Dutoi, T.R. Furlani, S.R. Gwaltney, A.~Heyden, S.~Hirata, C.-P. Hsu,
  G.~Kedziora, R.Z. Khalliulin, P.~Klunzinger, A.M. Lee, M.S. Lee, W.~Liang,
  I.~Lotan, N.~Nair, B.~Peters, E.I. Proynov, P.A. Pieniazek, Y.M. Rhee,
  J.~Ritchie, E.~Rosta, C.D. Sherrill, A.C. Simmonett, J.E. Subotnik, H.L.
  Woodcock, W.~Zhang, A.T. Bell, A.K. Chakraborty, D.M. Chipman, F.J. Keil,
  A.~Warshel, W.J. Hehre, H.F. Schaefer, J.~Kong, A.I. Krylov, P.M.W.
  Gill, M.~Head-Gordon, 
Phys. Chem. Chem. Phys. 8~(27) (2006) 3172.

\bibitem{May2004}
V.~May, O.~K\"uhn, {Charge and Energy Transfer Dynamics in Molecular Systems},
  WILEY-VCH Verlag GmbH \& Co.KGaA, 2004.

\bibitem{Egorov1999}
S.A. Egorov, K.F. Everitt, J.L. Skinner, 
J. Phys. Chem. A 103~(47) (1999) 9494.

\bibitem{Skinner2001}
J.L. Skinner, K.~Park, 
J. Phys.
  Chem. B 105~(28) (2001) 6716.

\bibitem{Stock2009}
G.~Stock, 
Phys.
  Rev. Lett. 102~(11) (2009) 1.

\bibitem{Tully1990}
J.C. Tully, 
J. Chem. Phys.
  93~(2) (1990) 1061.

\bibitem{Ben-Nun2000}
M.~Ben-Nun, J.~Quenneville, T.J. Mart\'{\i}nez, 
  J. Phys. Chem. A 104~(22) (2000) 5161.

\bibitem{Wu2005}
Y.~Wu, M.F. Herman, 
J. Chem. Phys. 123~(14) (2005) 144106.

\bibitem{cheng}
Y.C. Cheng, G.R. Fleming, 
{J. Phys. Chem. A} {112} ({2008}) {4254}.

\bibitem{mukamel}
S.~Mukamel, Principles of Nonlinear Optical Spectroscopy, Oxford, 1995.

\bibitem{nielsenchuang}
I.L. Chuang, M.A. Nielsen, 
{J. Mod. Opt.} {44} ({1997})
  {2455}.

\bibitem{cirac}
J.F. Poyatos, J.I. Cirac, P.~Zoller, 
Phys. Rev. Lett. 78~(2) (1997) 390.

\bibitem{YuenMohseni}
J.~{Yuen-Zhou}, M.~{Mohseni}, A.~{Aspuru-Guzik}, 
unpublished.
\urlprefix\url{http://arxiv.org/abs/1006.4866}.

\bibitem{sudarshan}
E.C.G. Sudarshan, P.M. Mathews, J.~Rau, 
Phys. Rev. 121~(3) (1961) 920.

\bibitem{minhaengbook}
M.~Cho, Two Dimensional Optical Spectroscopy, CRC Press, 2009.

\bibitem{goldmanbook}
M.~Goldman, Quantum Description of High-Resolution NMR in Liquids, Oxford
  University Press, 1991.

\bibitem{ginsberg}
N.S. Ginsberg, Y.-C. Cheng, G.R. Fleming, 
{Acc. Chem. Res.} {42} ({2009})
  {1352}.

\bibitem{jonas}
D.M. Jonas, 
Ann. Rev. Phys. Chem. 54
  (2003) 425.

\bibitem{minhaeng}
M.~Cho, 
{Chem. Rev.} {108}
  (2008) {1331}.

\end{thebibliography}

\end{document}